\documentclass[journal]{IEEEtran}
\normalsize
\usepackage{amsfonts, amsmath, amssymb, graphicx, caption, subfigure, setspace}
\usepackage{pstricks, pst-sigsys, pst-optexp, pst-node, pst-plot, pst-circ, moredefs}
\usepackage[colorlinks=true,linkcolor=blue,citecolor=blue,filecolor=blue,urlcolor=blue]{hyperref}
\usepackage{srcltx}

\begin{document}

\title{DSP Based PMD Emulators for Built-in Testing of Coherent Optical Receivers}

\author{A.~Gokcen~Mahmutoglu, Alper~T.~Erdogan and Alper Demir%
	\thanks{The authors are with the Department of Electrical and Electronics Engineering, Koc University, Rumeli Feneri Yolu, 34450 Sariyer-Istanbul, Turkey. (e-mail: \href{mailto:amahmutoglu@ku.edu.tr}{amahmutoglu@ku.edu.tr}; \href{mailto:alperdogan@ku.edu.tr}{alperdogan@ku.edu.tr}; \href{mailto:aldemir@ku.edu.tr}{aldemir@ku.edu.tr}).}%
	 }


\maketitle

\begin{abstract}
We propose discrete-time polarization mode dispersion (PMD) models that are compatible with the emerging coherent receiver techniques, and statistical sampling schemes for the model parameters. 
These models use multiple-input multiple-output (MIMO) finite impulse response (FIR) filters that are lossless and therefore lend themselves as perfect candidates for emulation of fiber channels suffering from PMD without polarization dependent loss (PDL). 
The concatenated composition of these filters resembles the continuous time lumped model of PMD channels and offers a flexible emulator and compensator structure in terms of computational complexity which constitutes the main bottleneck for real-time DSP applications.

The parameter sampling problem for accurate approximation of PMD channels considering their statistical behavior is tackled using three different approaches, which we introduce in order of decreasing deviation from the desired statistics and increasing computational complexity.
We present simulation results for each sampling method and compare them with the desired statistics.
\end{abstract}



\section{Introduction}
Polarization mode dispersion (PMD) is an important concern in the design of high-speed optical fiber communication systems.
Although recent advances in coherent receiver technologies \cite{ip2007digital,erdogan2008automatic,oktem2009adaptive} brought some relief from optical distortions in optical fibers with the use of compensation algorithms, PMD continues to be a major impediment at bit/symbol rates of 100 Gb/s and more \cite{savory2008digital,spinnler2010equalizer}.
PMD results in a random coupling and a speed difference between the two, normally degenerate, orthogonal polarization modes of propagation in single-mode optical fibers due to random imperfections breaking their circular symmetry, and therefore has a statistical nature \cite{poole1986phenomenological}. These effects result in pulse broadening and intersymbol interference (ISI) as well as the mixing of two orthogonal polarization channels which limit the transmission speed while increasing the system outage probability.

Accurate emulation of PMD is required for the development and testing of optical fiber communication systems. 
Currently, development of such systems is based on testing procedures employing conventional PMD emulators built with optical and mechanical components. 
These emulators can be constructed with different techniques such as simple polarization beam splitter (PBS) and phase retardation elements \cite{vgenis2010nonsingular} which only emulate first-order PMD, transmission loops with computer operated polarization controllers and polarization maintaining fibers (PMF) \cite{renaudier2008linear} and a concatenation of a high number of birefringent wave plates \cite{xie2010polarization}. 
The common property of these methods is that they either have limited emulation capability and programmability or suffer from high structural and computational complexity. 
Moreover, the incorporation of optical elements such as polarization scramblers and rotational mechanical systems results in expensive devices that are also bulky in size.

State of the art transceiver technologies make use of coherent receiver techniques and digital signal processing (DSP) algorithms in order to increase spectral efficiency and system capacity and compensate for the shortcomings of the legacy technology such as low tolerance for PMD at data rates higher than 40 Gb/s \cite{borne2005pmd}. 
Furthermore, as counterparts of PMD emulators with optical components, PMD compensators used in direct detection receiver schemes suffer from the same inconveniences of complexity and bulkiness.
In this paper, we propose PMD emulation methods based on a discrete-time scheme that is compatible with the emerging coherent receiver techniques.
We develop algorithms that can be used on custom chips already present in high speed coherent receivers with minimal alterations of these architectures or on ordinary digital processors for impairment assessment methods relying on off-line data processing. 

This alternative DSP based approach enables the development of more compact and flexible emulators to replace the existing analog and optical counterparts. 
The programmability feature of these emulators enables the coverage of a large span of fiber line scenarios via a simple modification of algorithm parameters.
The proposed scheme uses multiple-input multiple-output (MIMO) FIR filters that are lossless and therefore lend themselves as perfect candidates for emulation of fiber channels suffering from PMD without polarization dependent loss (PDL). 
The concatenated composition of these filters resembles the continuous-time lumped model of PMD channels and offers a flexible emulator structure in terms of computational complexity which constitutes the main bottleneck for real-time DSP applications.

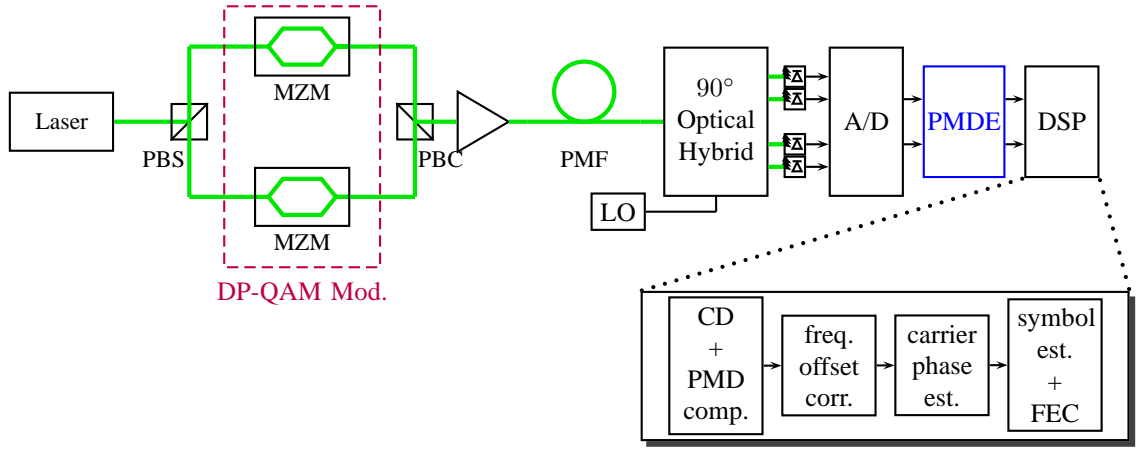
\begin{figure*}
  \begin{center}
    \begin{pspicture}[showgrid=false](0,-2)(13.5,4)
  \addtopsstyle{Fiber}{linecolor=green!90!black,linewidth=2\pslinewidth}
  \psset{labeloffset=0.6}
  \psset{usefiberstyle}
  \pnode(0,2.5){laser}
  \pnode([Xnodesep=1]laser){pbs}
  \pnode([offset=1]pbs){mzm1in}
  \pnode([offset=-1]pbs){mzm2in}
  \pnode([Xnodesep=3]mzm1in){mzm1out}
  \pnode([Xnodesep=3]mzm2in){mzm2out}
  \pnode([offset=-1]mzm1out){pbc}
  \pnode([Xnodesep=1.5]pbc){txout}

  \optbox[endbox,labeloffset=0, fiber](pbs)(laser){Laser}
  \beamsplitter[labelangle=-125,bssize=0.5](laser)(pbs)(mzm1in){PBS}
  \beamsplitter[labelangle=35,bssize=0.5](mzm1out)(pbc)(txout){PBC}
  \optmzm(mzm1in)(mzm1out){MZM}
  \optmzm(mzm2in)(mzm2out){MZM}
  \optamp[position=0.6](pbc)(txout){}
  \psline[style=Fiber](mzm2in)(pbs)(mzm1in)
  \psline[style=Fiber](mzm2out)(pbc)(mzm1out)

  \fnode[style=Dash,linecolor=purple,framesize=2.1 3.5]([Xnodesep=1.5,offset=-0.2]pbs){box}
  \nput{270}{box}{\textcolor{purple}{DP-QAM Mod.}}
  
  \pnode([Xnodesep=1.5]txout){rxin}
  \pnode([offset=0]rxin){hybrin1}
  \psblock([Xnodesep=-0.3,offset=-1.2]rxin){lo}{LO}
  \pnode([Xnodesep=0.3]hybrin1){hybrin}
  \pnode([Xnodesep=1.3]lo){loin}

  \psline[style=Fiber](rxin)(hybrin1)(hybrin)
  \optfiber[fiberloops=1, labelalign=b](txout)(rxin){PMF}
  
  \psfblock[framesize=1.4 2]([Xnodesep=1]rxin){hybr}{\parbox[c]{1.1\psunit}%
  {\centering $90^{\circ}$ \\ Optical \\ Hybrid}}

  \nclist{-}{ncline}{lo,loin,hybr}
  
  \pnode([offset=0.6, Xnodesep=0.9]hybr){det1}
  \pnode([offset=0.3, Xnodesep=0.9]hybr){det2}
  \pnode([offset=-0.3, Xnodesep=0.9]hybr){det3}
  \pnode([offset=-0.6, Xnodesep=0.9]hybr){det4}
  
  \optdetector[fiber, detsize=0.3,dettype=diode]([offset=0.6]hybr)(det1){}
  \optdetector[fiber, detsize=0.3,dettype=diode]([offset=0.3]hybr)(det2){}
  \optdetector[fiber, detsize=0.3,dettype=diode]([offset=-0.3]hybr)(det3){}
  \optdetector[fiber, detsize=0.3,dettype=diode]([offset=-0.6]hybr)(det4){}
  
  \psfblock[framesize=1 2]([Xnodesep=2]hybr){adconv}{\parbox[c]{1.1\psunit}{\centering A/D}}
   
  \pnode([Xnodesep=1.2]hybr){hybrout}

  \psset{arrows=->}
  \ncline[offset=0.6]{hybrout}{adconv}
  \ncline[offset=0.3]{hybrout}{adconv}
  \ncline[offset=-0.3]{hybrout}{adconv}
  \ncline[offset=-0.6]{hybrout}{adconv}
  
  \psfblock[framesize=1.1 1.5,linecolor=blue]([Xnodesep=1.3]adconv){pmde}{\parbox[c]{1.1\psunit}{\centering \textcolor{blue}{PMDE}}}
  \ncline[offset=-0.3]{adconv}{pmde}
  \ncline[offset=0.3]{adconv}{pmde}

  \psfblock[framesize=1 1.5]([Xnodesep=1.3]pmde){dsp}{\parbox[c]{1.1\psunit}{\centering DSP}}
  \ncline[offset=-0.3]{pmde}{dsp}
  \ncline[offset=0.3]{pmde}{dsp}

  \pnode([Xnodesep=-0.5,offset=-0.75]dsp){dsplcor}
  \pnode([Xnodesep=0.5,offset=-0.75]dsp){dsprcor}
  \pnode([Xnodesep=-5.1,offset=-1.5]dsplcor){dspzoom1}
  \pnode([Xnodesep=0.4,offset=-1.5]dsprcor){dspzoom2}

  \psline[arrows=-,linestyle=dotted,linewidth=2\pslinewidth](dsplcor)(dspzoom1)
  \psline[arrows=-,linestyle=dotted,linewidth=2\pslinewidth](dsprcor)(dspzoom2)
  \psframe[shadow=true]([offset=-2]dspzoom1)(dspzoom2)

  \psblock([Xnodesep=1,offset=-1]dspzoom1){pmdcomp}{\parbox[c]{1\psunit}{\centering CD \\ + \\ PMD \\ comp.}}
  \psblock([Xnodesep=1.5]pmdcomp){focorr}{\parbox[c]{1\psunit}{\centering freq. \\offset \\ corr.}}
  \psblock([Xnodesep=1.5]focorr){cpest}{\parbox[c]{1\psunit}{\centering carrier \\ phase \\ est.}}
  \psblock([Xnodesep=1.5]cpest){symest}{\parbox[c]{1\psunit}{\centering symbol \\ est. \\ + \\ FEC}}
  \nclist{ncline}{cdcomp,pmdcomp,focorr,cpest,symest}

\end{pspicture}
  \end{center}
  \caption{Outline of an optical communication system with a coherent receiver and the addition of the proposed PMD emulator (PMDE). PBS: polarization beam splitter, MZM: Mach-Zehnder modulator, PBC: polarization beam combiner, LO: local oscillator, A/D: analog to digital converter, DSP: digital signal processor, CD: chromatic dispersion, FEC: forward error recovery \cite{savory2008digital,renaudier2008linear,nelson2009performance,ludwig2011ultrafast}.}
  \label{fig:tranceiver}
\end{figure*}

Without a restriction on the number of filter sections, the construction of discrete-time PMD emulators is straightforward. As it will be demonstrated in Section \ref{sec:UPD}, with a growing number of filter sections, the properties of the PMD emulator converge to those of a real fiber channel (similar to continuous-time models with a large number of birefringent sections).
However, because a discrete-time filter with hundreds of sections would be  of no practical interest, we investigate methods for approximating the channel properties of a real fiber with a computationally tractable and implementationwise feasible number of filter sections.
Therefore, the real challenge in employing FIR MIMO filters for PMD emulation presents itself as a statistical parameter sampling problem.
We strive to generate a statistical ensemble of filter parameters that produce  the correct stochastic behavior of PMD.
In the literature, the accuracy of PMD models is usually quantified in terms of the so-called \emph{order} of PMD.
The definition of PMD orders is based on the Taylor expansion of the PMD vector around a frequency point \cite{foschini1991statistical}. Since the PMD vector has a statistical nature, the individual terms of this expansion have also a random character.
However, low order PMD models take only the first few terms of this expansion into account and therefore cannot describe the complex frequency dependent behavior of PMD. For instance, a first-order PMD model uses only the constant term of the Taylor expansion and hence can only describe frequency independent effects. Therefore, these models are only valid for narrowband signals and become increasingly inaccurate as the frequency range of interest grows \cite{chen2003bandwidth}.
In our treatment, we make no assumption on the bandwidth of the signal and take also the frequency dependent behavior of PMD into account. Our methods do not rely on a Taylor expansion around a single frequency. On the contrary, we consider the joint statistical properties of PMD vectors at different frequencies.
We strive to produce the correct PMD statistics over the whole frequency range of interest and have the correct frequency autocorrelation \cite{karlsson1999autocorrelation,shtaif2000study}. This way we build a PMD emulator that can also be used for the testing of wideband systems.

In the next section, we give a brief outline of the paper and describe the sampling problem at hand without going into details.
Sections \ref{sec:dt_ll_sys} and \ref{sec:UPD} are dedicated to the presentation of paraunitary MIMO FIR filters and their properties regarding PMD emulation.
In order to address the filter parameter sampling problem, we propose three different techniques and introduce them in Section \ref{sec:param_sample} in order of decreasing deviation from the desired statistics and increasing computational complexity. 
Finally, we conclude with simulation results and a brief summary.

\section{From Analog to Digital PMD Emulation}

Figure \ref{fig:tranceiver} depicts the key components of an optical communication system with a coherent receiver \cite{savory2008digital,renaudier2008linear,nelson2009performance,ludwig2011ultrafast}.
Note that this back-to-back configuration of the transmitter and the receiver lacks the transmission fiber of a real communication link. 
A conventional PMD emulator apparatus, usually present between the receiver and the transmitter for testing purposes, is also absent. 
Instead, the set-up includes a polarization maintaining fiber to connect the transceiver and the receiver. 
The distortion of the signal for PMD emulation purposes is performed after it is converted to the electrical domain with optical detectors and then sampled with the analog-to-digital converter (A/D). 
At this stage, the sampled and digitized signal consists of two complex valued components, one for each polarization. 
In recent high speed experiments with up to 10 Tb/s for a single channel, both of these polarization components at the receiver are modulated using 16-QAM independently \cite{richter2011single,palushani2011transmission}, which corresponds to a dual polarization quadrature amplitude modulation (DP-QAM) with a 16-ary symbol constellation. 
Usually, after the A/D conversion stage, the signal is treated with a DSP, compensating for chromatic dispersion (CD) and PMD, and put through a decision circuit after frequency offset correction and carrier phase estimation.

The PMD compensation can be handled by different schemes based on DSP algorithms depending on the requirements of the channel, the modulation method and the transmission speed. 
For channels, where the pulse broadening due to PMD, i.e. ISI, is negligible, adaptive techniques such as the constant modulus algorithm (CMA) \cite{godard1980self} are used for the demultiplexing of the two polarization components \cite{savory2008digital,renaudier2008linear,xie2010polarization}. 
For this task, ``a blind source separation scheme based on the magnitude boundedness'' of the signals was also proposed in \cite{oktem2009adaptive}. 
For higher data rates, an additional compensation of the polarization components needs to be performed in time as well. 
This can be achieved with methods such as a modification of the CMA algorithm, a fractionally spaced equalizer (FSE) \cite{ip2007digital,kaneda2009coherent}, and algorithms exploiting the paraunitary structure of the PMD channel \cite{oktem2010adaptive}. In order to evaluate the performance of a PMD compensation method, real life tests are conducted either with actual fibers of realistic lengths \cite{savory2008digital,fischer2011high} or with PMD emulators \cite{vgenis2010nonsingular,renaudier2008linear,xie2010polarization}. 
Following the example of PMD compensation methods, in order to exploit the advantages of a DSP based emulation technique, we propose to insert a PMD emulation stage between the A/D and the DSP in Figure \ref{fig:tranceiver} considering the fact that the signal is already sampled for further processing.

A 2-input 2-output discrete-time filter is required to emulate the effects of a real PMD channel which are of statistical nature. That means we have to choose the filter parameters according to some statistical law that results in accurate channel emulation characteristics such as mean DGD value as well as the whole probability distribution of the DGD and frequency dependent statistics determined by the autocorrelation function (ACF) of the PMD vector \cite{khosravani2001time}. 
We can use these quantities as performance criteria despite having control only on the parameters of the filter, because we can compute them if we know the transfer function of the filter.
Hence, the discrete-time model of a PMD channel can be thought as a generator of PMD vectors with adjustable parameters. 
For each parameter set, we get a different PMD vector and DGD value at the output.

Similar to the analog case \cite{demir2008emulation}, we can construct inherently lossless discrete-time filters for PMD emulation. 
Each such filter can be decomposed into individual one-tap filter sections (degree-one filters) with DGD values of one symbol period, and conversely, one can construct filters by concatenation of these sections with an additional degree of freedom for the filter length. 
Although all the sections have the same DGD value of one symbol period, the principal states of polarization (PSP) determined by a single $2\times 1$ complex vector parameter for each section is different, and this causes a single state of polarization to propagate at different speeds in each section. 
The orientations of the PSPs with respect to each other determine the resulting distortion on the signal. 
Choosing these parameters randomly and in an appropriate manner, we can capture the statistical nature of PMD correctly. 
This is posed as a problem of random parameter sampling for lossless filters.

Later in Section \ref{sec:param_sample} we will propose three methods for building discrete time paraunitary models achieving the desired statistical properties. 
The first parameter sampling scheme, Cascaded Sampling Method (CSM), offers an answer to the following question. 
How should we sample the filter parameters such that an ensemble of discrete-time filters has an adjustable mean DGD value that is constant over the frequency range of interest? 
This first step in constructing a discrete-time filter for PMD emulation is important because, as it will be demonstrated in Section \ref{sec:UPD}, when their parameters are not sampled in a special manner, i.e. when they are sampled uniformly and independently, the DGD distribution of lossless filters depends only on the number of sections with a corresponding mean DGD value that grows linearly with the square root of this number. 
Hence for a fixed filter length and i.i.d. uniform filter parameters, the DGD is not adjustable. Moreover, for filters with a computationally feasible number of sections, the probability density function (PDF) of the DGD deviates heavily from the desired Maxwellian PDF, especially in the tail region. 
This deviation is of great importance for the testing of communication systems because the tail of the PDF describes the relative frequency of high DGD values and it is these values that in turn determine the system outage probabilities. 
Consequently, we want to construct a statistical model parameter sampling scheme that produces Maxwellian DGD PDFs with adjustable means. 
Another important property we expect is that these statistics are constant over the whole frequency range of interest.

Although we strive for a DGD distribution that is stationary over the frequency range of interest, it proves useful to concentrate on its behavior at the center frequency. 
It turns out that an instantaneous DGD value  at the center frequency has a simple relationship with the filter parameters. 
This relationship boils down to a sum of random variables problem. 
Instead of considering the $N$ filter sections separately and trying to sample their parameters individually, we think of the whole filter as a composition of 2-section blocks. 
This way we construct building blocks that no longer have a constant DGD. 
The merging of two degree-one filter sections results in an elementary building block which can also produce fractional DGD values and hence has more expression capacity. 
This leaves us a greater elbowroom for the design of full length filters in the time domain and sample their parameters accordingly. 
The parametrization of two-section blocks is, however, different than their one-section counterparts, and since the implementation of paraunitary filters requires the knowledge of the parameters of individual degree-one filter sections, a mapping from the two-section block parametrization to the usual one-section parametrization remains to be found. 
In the proposed CSM, this mapping is translated into an eigenvalue problem to give the degree-one filter section parameters directly without the need for the costly calculation of transfer functions, resulting in a computationally efficient parameter sampling scheme.



The second method we propose is based on an indirect technique for sampling the filter parameters. 
This method, Compensated Markov Chain Monte Carlo (C-MCMC), is based on a modified version of the standard Metropolis-Hastings algorithm \cite{metropolis1953equation,hastings1970monte}. 
This algorithm devises a random walk in the space of filter parameters and the generated samples are the values it assumes during the course of this walk. 
The random walk can be conceived as a search with an objective or reward function but without a definite target point such as a maximum or a minimum. 
There are more rewarding, hence more desired points in the filter parameter space since these points produce more probable PMD vector values. At the end, the probability distribution of the ensemble of points visited during the random walk converges to the desired PDF. The random jumps from point A to the point B occur with the help of a decision rule that reflects the weighting of the desired PDF over the space of PMD vectors.

Contrary to the CSM, C-MCMC uses the whole information contained in the PDFs of PMD vectors and not just their Euclidean norms. This enables us to capture the autocorrelation function of PMD vectors over the frequency range of interest at the expense of increased computational complexity.

In the compensated MCMC scheme, the random walk is constructed in a different space than the decision rule operates in.
The decision rule is  based on a ratio of PDF values of PMD vectors but we cannot sample these directly. Each set of filter parameters results in a PMD vector at the output but this mapping is neither one-to-one nor isometric.
Hence, sampling filter parameters by looking at the corresponding PMD vectors can fail if for example a large number of parameter sets are mapped to the same PMD vector.

The last method we propose is different from the previous two in that it is not probabilistic in nature. 
Instead we employ a greedy transfer function approximation method (GTFA) \cite{tkacenko2006iterative} in order to capture not only the PMD vector statistics but also the whole transfer function behavior of the channel. 
We start with the transfer function of a PMD channel generated with a full model. 
A full model in this context means a PMD channel model with a high number (several hundred to one thousand) of concatenated birefringent sections. 
Such models are known to accurately approximate PMD channels when their parameters are sampled uniformly and independently.
After generating a full model and computing its transfer function, we generate a discrete-time model with a low number of sections and proceed with the greedy approximation. 
This algorithm optimizes each section of the discrete-time filter locally such that the whole filter with other sections held fixed matches the full model according to some closeness measure. 
The first part of the local approximation boils down to the closest unitary matrix problem which is easily solved by a singular value decomposition based algorithm.
This elegant solution is followed by the second part which uses an eigenvalue equation to finalize the approximation. 
This procedure is repeated until the desired level of match is obtained. 
In order to generate another set of filter parameters, this whole routine is repeated. 
As expected, this method achieves the best results in terms of statistical accuracy of the generated models but it is also the most computationally complex of the three.

All three of these parameter sampling schemes for paraunitary filters are designed to match their PMD vector and DGD statistics to the statistics of a real optical fiber. 
The goodness of match is measured in terms of the PDFs and their joint moments. 
A perfect match, in this regard, corresponds to all-order PMD emulation. 
A partial match, such as in the case of the CSM, represents only the lower orders of PMD. 
However, one point worthy of notice is that there is no direct connection between the order of moments of the joint PMD distribution on the frequency axis and the order of PMD emulation in the sense it is commonly used in the literature. 
For example, emulating first-order PMD corresponds to generating instantaneous DGD values that are constant over the frequency range of interest and distributed with a Maxwellian PDF in time. 
The frequency dependent behavior is captured by higher order PMD. 
Since all three of the methods we propose produce the required marginal PDFs at every frequency, we are interested in the relationship of our target values at different frequencies. 
That relationship is given by the statistics of the joint PDFs such as the ACF on the frequency axis. 
Therefore, matching the behavior of the ACF results in all-order PMD emulation. 
On the other hand, a constant ACF amounts to emulating only the first-order PMD. 
Everything in-between these extreme points emulates higher order PMD to an extent depending on how good the ACF match is. In this sense, all three parameter sampling schemes we propose take all orders of PMD into account and emulate them with a trade-off between accuracy and computational complexity.

\section{Discrete Time Lossless Systems - the Basics}
\label{sec:dt_ll_sys}
A matrix transfer function $\mathbf{H}(z)$ is said to be paraunitary if the following holds \cite{vaidyanathan1989role}:

\begin{equation}
  \mathbf{H}^*(z^{-*})\mathbf{H}(z) = c^2\mathbf{I}_r \quad \forall z.
  \label{eqn:def_paraunitary}
\end{equation}

\noindent Here,  the superscript ``*'' denotes conjugation and transposition, $\mathbf{I}_r$ is the $r \times r$ identity matrix and $c$ is a scalar constant.

If all the entries of $\mathbf{H}$ are stable transfer functions and equation (\ref{eqn:def_paraunitary}) is satisfied with $c=1$, then $\mathbf{H}$ is unitary on the unit circle and therefore corresponds to a lossless LTI system. 
Conversely, it can be shown that for rational transfer functions, the unitariness of $\mathbf{H}(e^{j\omega})$ for all $\omega$ implies (\ref{eqn:def_paraunitary}) with $c=1$.
Hence a lossless system can be defined as a causal, stable paraunitary system \cite{vaidyanathan1993multirate}. Since FIR filters are inherently stable, from now on the terms ``lossless'' and ``paraunitary'' will be used interchangeably for causal systems.

A general $M \times M$ degree-one FIR lossless transfer function can be written as \cite{vaidyanathan1993multirate},

\begin{equation}
  \mathbf{H}_{i}(z) = [\mathbf{I} - \mathbf{v}_{i}\mathbf{v}^{*}_{i} + z^{-1}\mathbf{v}_{i}\mathbf{v}^{*}_{i}]\mathbf{H}_i(1) \quad.
  \label{eqn:ll_degree1}
\end{equation}

\noindent Here, $\mathbf{v}_i$ is a complex $M \times 1$ vector of unit norm and $\mathbf{H}_i(1)$ is a unitary matrix. It can easily be verified that this transfer function is paraunitary. The extension to higher degree systems follows naturally with the concatenation of such degree-one blocks and a multiplication with a constant unitary matrix $\mathbf{R}$.

\begin{equation}
  \mathbf{H}(z) = \mathbf{H}_{1}(z) \mathbf{H}_{2}(z) \dots \mathbf{H}_{N}(z)\mathbf{R}
  \label{eqn:ll_degreeN}
\end{equation}
Note that every lossless transfer function can be expressed in this form and this factorization is not unique.

In PMD related calculations we will be dealing with $2 \times 2$ paraunitary transfer functions that act on $2 \times 1$ vector valued time series consisting of two orthogonal polarization components of the transmitted signal. In this regard the expression of the corresponding DGD can be obtained based on the expression in (\ref{eqn:ll_degreeN}). The distance between the fast and slow principal states of polarization can be expressed as the difference \cite{damask2005polarization}

\[
\tau(\omega) = |\Im\{\lambda_1[\mathbf{G}(e^{j\omega})] - \lambda_2[\mathbf{G}(e^{j\omega})]\}| \quad,
\]

\noindent where $\Im\{\lambda_i[\mathbf{G}(e^{j\omega})]\}$ denotes the imaginary part of the $\mathrm{i^{th}}$ eigenvalue of $\mathbf{G}(e^{j\omega})$ and,

\begin{equation}
  \mathbf{G}(e^{j\omega}) = \frac{\mathrm{d}\mathbf{H}(e^{j\omega})}{\mathrm{d}\omega}\mathbf{H}^{*}(e^{j\omega}). 
  \label{eqn:matrix_G}
\end{equation}

\noindent If we now substitute equation (\ref{eqn:ll_degreeN}) in (\ref{eqn:matrix_G}) we get

\begin{equation}
  \mathbf{G}(e^{j\omega}) = -j\sum \limits_{i=1}^{N} \prod \limits_{j=1}^{i} \mathbf{H}_{j-1}\mathbf{v}_{i}\mathbf{v}_{i}^{*}\mathbf{H}_{j-1}^{*} \quad,
  \label{eqn:matrix_G_prod}
\end{equation}

\noindent with $\mathbf{H}_0 = \mathbf{I}_2$. 
This expression can be greatly simplified for $\omega = 0$ if we set $\mathbf{H}_{k}(1) = \mathbf{I}_2$ for all $k$. 
The DGD of the fiber, $\tau(\omega)$, evaluated at the center frequency can in this case be computed to be the difference between the two eigenvalues of the positive definite matrix $\mathbf{V} \mathbf{V}^*$ with $\mathbf{V} = [\mathbf{v}_{1} \mathbf{v}_{2} \dots \mathbf{v}_{N}]$. 
Calculating the eigenvalues of this matrix results in a quadratic equation, and twice the discriminant of this equation gives us the difference between the two eigenvalues. 

\begin{equation}
  \tau(0) = \sqrt{(a - b)^2 + 4cc^*}
  \label{eqn:tau_zero}
\end{equation}

\noindent where,

\begin{equation}
  j\mathbf{G}(1) = 
  \left[
  \begin{array}[pos]{cc}
    a & c \\
    c^* & b 
  \end{array}
  \right], \;
  \mathbf{v}_i = 
  \left[
  \begin{array}[pos]{c}
    \alpha_i \\
    \beta_i 
  \end{array}
  \right], \;
  \begin{array}{l}
    a = \sum_{i=1}^N |\alpha_i|^2 \\
    b = \sum_{i=1}^N |\beta_i|^2 \\
    c = \sum_{i=1}^N \alpha_i \beta_i^*
  \end{array}
  \label{eqn:matrix_G_structure}
\end{equation}

\section{DGD Statistics of Paraunitary Filters with Uniform i.i.d. Parameters}
\label{sec:UPD}

As a reference point and motivation for further discussion, we investigate the statistical behavior of an ensemble of paraunitary filters when their parameters are sampled uniformly and independently. 
As mentioned in the introduction, the structure of paraunitary FIR filters as concatenation of degree-one building blocks is similar to the continuous time lumped model of optical fibers \cite{demir2008emulation} and their DGD distribution therefore resembles a real fiber. 
This resemblance is, however, limited with two important hurdles: The mean DGD of filters tailored with this selection method is fully determined only by the number of degree-one sections, $N$, and is not adjustable. Adjustability is a crucial feature for a PMD emulator. 
Furthermore, the probability density functions deviate heavily from the desired curves particularly in the tails which correspond to high DGD cases and are therefore of primary importance for system outage probability considerations.

\begin{figure}
	\centering
		\input{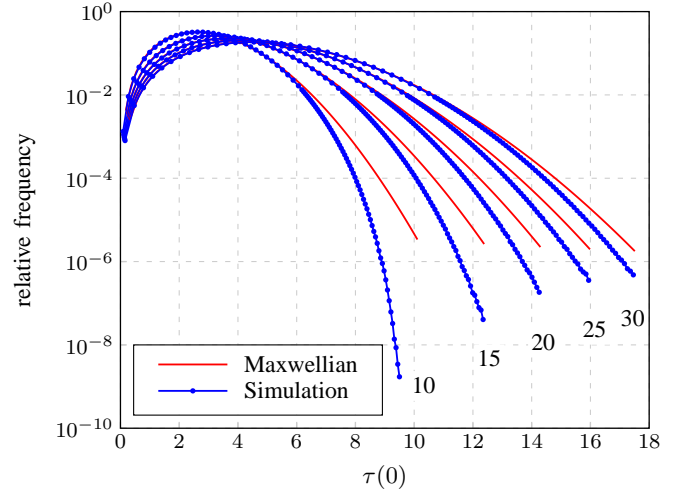}	
	\caption{Probability density functions of the DGD for various number of sections (numbers at the end of the curves) for i.i.d.\ uniform filter parameters.}
\label{fig:uni_DGD_pdf}
\end{figure}

The results for the DGD distribution are given in Figure \ref{fig:uni_DGD_pdf} for the uniform i.i.d.\ parameter sampling method:
\[
\alpha_i =  \sqrt{X_i} e^{j2\pi Y_i} \qquad \beta_i = \sqrt{1 - |\alpha_i|^2} = \sqrt{1-X_i} \quad,
\]
\noindent where $X_i$ and $Y_i$ are uniformly distributed in $[0 ,1]$. The $X_i$s constitute a set of $N$ independent random variables. The same applies for $Y_i$ and these two sets are also independent from each other. The distribution of the DGD has been computed for  different number of sections, $N=10$ to $N=30$ with $10^9$ samples and compared with the expected Maxwellian distributions with the same mean as the generated data set. Although the PDF deviation becomes less prominent as $N$ increases, we cannot rely on this behavior for statistical accuracy since the required number of sections and hence the computational complexity of the resulting emulator would be too high.

Together with equation (\ref{eqn:tau_zero}), (\ref{eqn:matrix_G_structure}) and the parameter selection method above, the expression for the DGD becomes

\[
\tau = \sqrt{(a - b)^2 + 4cc^*} 
\]
\begin{equation}
  = \sqrt{(2a - N)^2 + (2\Re\{c\})^2 + (2\Im\{c\})^2}
  \label{eqn:tau_of_G}
\end{equation}

\begin{align*}
=  \Bigg[\left(2\sum_{i=1}^{N} X_i - N\right)^2 &+  \left(2\sum_{i=1}^{N} \sqrt{X_i-X_i^2}\cos(2\pi Y_i)\right)^2  \\
  &+  \left(2\sum_{i=1}^{N} \sqrt{X_i-X_i^2}\sin(2\pi Y_i)\right)^2 \Bigg]^\frac{1}{2} 
\end{align*}

Remembering that the Maxwellian distribution can be expressed as the square root of sum of squares of three independent zero-mean Gaussian random variables with the same variance, the above expression can be used to verify that the distribution of $\tau$ indeed approximates a Maxwellian distribution. 
This is achieved invoking the central limit theorem on each of the three summations of length $N$.
The first term, $2\sum_{i=1}^{N} X_i - N$, is zero-mean. Because $X_i$ are independent, the total random variable $2a-N$ has the sum of variances of $X_i$ as its variance. 

\[
\mathrm{var}(2a-N) = 4N \mathrm{var}(X_i - \frac{1}{2}) = \frac{N}{3}
\]

It is again straightforward to calculate the mean and the variance of the second and the third term of the sum.
Because $X_i$ are independent among each other and also independent of $Y_i$ one can see that $\Re\{c\}$ is again zero mean because of the symmetry of the PDF of $\cos(2\pi Y_i)$ around zero.
Furthermore the variance of $2\Re\{c\}$ can be computed as

\begin{align*}
\mathrm{var}(2\Re\{c\}) &= N\mathrm{var}\left(2\sqrt{X_i-X_i^2}\right)\mathrm{var}(\cos(2\pi Y_i)) \\
&= 4N\left(\frac{1}{2}-\frac{1}{3}\right)\frac{1}{2} = \frac{N}{3}\quad.
\end{align*}

The same applies for $2\Im\{c\}$ and hence the total random variable $\tau$ consists of three approximately normally distributed random variables with $\mu \approx 0$ and $\sigma^2 \approx N/3$. This enables us to compute the expected mean of $\tau$ as
\begin{equation}
  \bar{\tau} \approx \sqrt{\frac{8}{\pi}}\sigma = \sqrt{\frac{8}{3\pi}N}\quad.
  \label{eqn:indep_mean_tau}
\end{equation}

These results are valid for the center frequency, $\omega = 0$ but further investigation reveals that with uniform i.i.d.\ parameter selection they also hold at other frequencies since there is no bias for any specific frequency point.
More in depth discussion of innate statistical properties of PMD emulators with i.i.d.\ uniform parameters including the derivation of the exact DGD  distribution and the examination of the deviation from the Maxwellian is outside of the scope of this paper, but can be found in \cite{karlsson2001probability}.

\section{Parameter Sampling Methods for Accurate PMD Emulation}
\label{sec:param_sample}
The discussion in Section \ref{sec:UPD} suggests that a different parameter sampling scheme must be implemented in order to achieve accurate DGD statistics. Targeting the DGD statistics is only the first step in the construction of a PMD emulator, since joint PDF statistics for PMD vectors at different frequencies, that reflect the frequency dependent behavior, must be considered as well. To this end, in this section we propose three different parameter sampling methods for discrete time paraunitary FIR filters which can be used for more accurate PMD emulation. From such an emulator we expect a Maxwellian DGD distribution with an adjustable mean value and frequency independent behavior over the whole frequency range of interest. In order to capture higher order effects,  we also require a good approximation of the frequency autocorrelation of the PMD vector \cite{khosravani2001time}. With respect to these performance criteria, the three random parameter sampling schemes we propose can be listed and classified as follows:

\begin{itemize}
  \item \emph{Cascaded Sampling Method (CSM)}: Accurate emulation of first-order PMD with limited control over higher order effects.
  \item \emph{Compensated Markov Chain Monte Carlo Method (C-MCMC)}: Matching the marginal PDFs and the ACF of PMD vectors. Accurate higher order PMD emulation.
  \item \emph{Greedy Transfer Function Approximation Method (GTFA)}: Transfer function matching and hence accurate all-order PMD emulation.
\end{itemize}

There is a direct link between this classification and the computational complexity of the sampling schemes: The least ``general'' method is also the fastest.  

Following sections introduce the aforementioned sampling schemes and demonstrate their performance emulating an optical communication link with a bandwidth of 40 GHz and a mean DGD of $0.4$ symbol period. All of the experiments use a discrete time filter consisting of 20 birefringent sections.  Furthermore we assume that the continuous time system was sampled four times over its minimum rate ($4 \times 40$ GHz) and therefore set the mean DGD of the emulators to $1.6$ sampling periods.

\subsection{Cascaded Sampling Method}
\label{sec:cascading}
In order to sample the filter parameters in such a way so that the mean DGD becomes adjustable and frequency independent, one can extend the degree-one paraunitary building blocks of the filter to 2-section blocks that have dependent $\mathbf{v}_i$. We call this technique ``cascaded sampling'' and analyze its properties below.

\subsubsection{Constructing a Paraunitary FIR Filter with Maxwellian DGD Distribution at the Center Frequency}

The difference between the eigenvalues of a Hermitian matrix,

\begin{equation}
  \mathbf{G} = \left[
	\begin{array}[pos]{cc}
		A & C \\
      		C^* & B 
	\end{array} 
	\right] \quad,
    \label{eqn:structure_G}
\end{equation}

\noindent has the form: $\tau = \sqrt{(A - B)^2 + 4CC^*}$. Therefore an ensemble of $2 \times 2$ Hermitian matrices will have a Maxwellian spacing distribution, i.e. the distribution of the difference of its eigenvalues, if the individual components in (\ref{eqn:structure_G}) are distributed as follows:

\begin{equation}
	\begin{array}{cc}
	  A \sim \mathcal{N}(\frac{N}{2},\sigma) & \Re{\{C\}} \sim  \mathcal{N}(0,\sigma) \\
	  B = N-A & \Im{\{C\}} \sim \mathcal{N}(0, \sigma)   
	\end{array}
  \label{eqn:components_G}
\end{equation}

\noindent Here $\mathcal{N}(\mu,\sigma)$ denotes the Gaussian distribution with mean $\mu$ and standard deviation $\sigma$. The resulting Maxwellian distribution will have the following mean value:

\begin{equation}
  \bar{\tau} = 2\sqrt{\frac{8}{\pi}}\sigma \quad.
  \label{eqn:mean_tau_of_sigma}
\end{equation}

At this stage one can ask if unit norm vectors $\mathbf{v}_i$ can be found such that $\mathbf{G} = \sum_{i=1}^N \mathbf{v}_i\mathbf{v}_i^*$. 
If one can find such vectors they can be used to construct a paraunitary FIR filter so that it will have a Maxwellian DGD at the center frequency by construction. 
This is indeed the case if $\mathrm{tr}(\mathbf{G}) \in \mathbb{Z}$ and $\mathrm{tr}(\mathbf{G}) \geq \mathrm{rank}(\mathbf{G})$ \cite{hartwig1990matrix}. 

Since $N = \mathrm{tr}(\mathbf{G})$ can be restraint to positive integers, this condition can always be satisfied.
Being the sum of rank 1 projection matrices, $\mathbf{v}_i\mathbf{v}_i^*$, puts one additional constraint on $\mathbf{G}$: $\mathbf{G}$ must be positive definite. This enforces the following inequalities on the parameters in (\ref{eqn:components_G}).

\begin{equation}
  A > 0, \quad B > 0, \quad AB > |C|^2
  \label{eqn:sigma_constraints}
\end{equation}

 Because of these constraints, the distributions of $A$ and $C$ in (\ref{eqn:components_G}) must have finite support and hence cannot be actual Gaussians.
 In Section \ref{DGD_constraints} we will show that adjusting the standard deviation of the distributions one can eliminate this discrepancy for all practical purposes.

\subsubsection{The case for $N=2$}
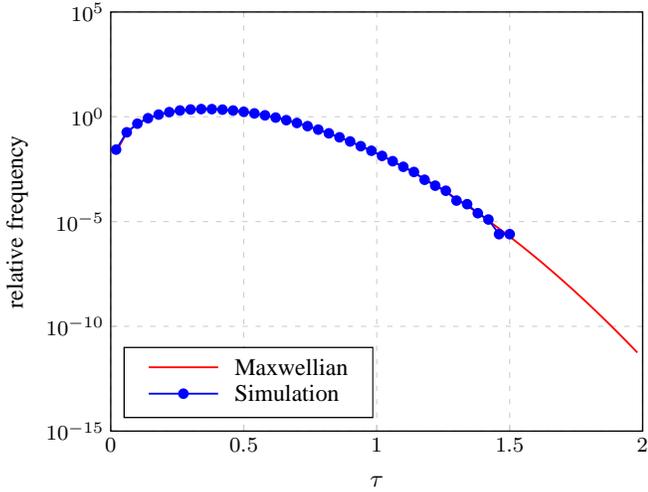
\begin{figure}
  \begin{center}
%
%
%
\providelength{\AxesLineWidth}       \setlength{\AxesLineWidth}{0.5pt}%
\providelength{\GridLineWidth}       \setlength{\GridLineWidth}{0.4pt}%
\providelength{\GridLineDotSep}      \setlength{\GridLineDotSep}{0.4pt}%
\providelength{\MinorGridLineWidth}  \setlength{\MinorGridLineWidth}{0.4pt}%
\providelength{\MinorGridLineDotSep} \setlength{\MinorGridLineDotSep}{0.8pt}%
\providelength{\plotwidth}           \setlength{\plotwidth}{8cm}
\providelength{\LineWidth}           \setlength{\LineWidth}{0.7pt}%
\providelength{\MarkerSize}          \setlength{\MarkerSize}{4pt}%
\newrgbcolor{GridColor}{0.8 0.8 0.8}%
%
\psset{xunit=0.441955\plotwidth,yunit=0.034857\plotwidth}%
\begin{pspicture}(-0.239631,-17.280702)(2.023041,5.526316)%


\psline[linestyle=dashed,dash=2pt 3pt,dotsep=\GridLineDotSep,linewidth=\GridLineWidth,linecolor=GridColor](0.000000,-15.000000)(0.000000,5.000000)
\psline[linestyle=dashed,dash=2pt 3pt,dotsep=\GridLineDotSep,linewidth=\GridLineWidth,linecolor=GridColor](0.500000,-15.000000)(0.500000,5.000000)
\psline[linestyle=dashed,dash=2pt 3pt,dotsep=\GridLineDotSep,linewidth=\GridLineWidth,linecolor=GridColor](1.000000,-15.000000)(1.000000,5.000000)
\psline[linestyle=dashed,dash=2pt 3pt,dotsep=\GridLineDotSep,linewidth=\GridLineWidth,linecolor=GridColor](1.500000,-15.000000)(1.500000,5.000000)
\psline[linestyle=dashed,dash=2pt 3pt,dotsep=\GridLineDotSep,linewidth=\GridLineWidth,linecolor=GridColor](2.000000,-15.000000)(2.000000,5.000000)
\psline[linestyle=dashed,dash=2pt 3pt,dotsep=\GridLineDotSep,linewidth=\GridLineWidth,linecolor=GridColor](0.000000,-15.000000)(2.000000,-15.000000)
\psline[linestyle=dashed,dash=2pt 3pt,dotsep=\GridLineDotSep,linewidth=\GridLineWidth,linecolor=GridColor](0.000000,-10.000000)(2.000000,-10.000000)
\psline[linestyle=dashed,dash=2pt 3pt,dotsep=\GridLineDotSep,linewidth=\GridLineWidth,linecolor=GridColor](0.000000,-5.000000)(2.000000,-5.000000)
\psline[linestyle=dashed,dash=2pt 3pt,dotsep=\GridLineDotSep,linewidth=\GridLineWidth,linecolor=GridColor](0.000000,0.000000)(2.000000,0.000000)
\psline[linestyle=dashed,dash=2pt 3pt,dotsep=\GridLineDotSep,linewidth=\GridLineWidth,linecolor=GridColor](0.000000,5.000000)(2.000000,5.000000)


\psline[linewidth=\AxesLineWidth,linecolor=GridColor](0.000000,-15.000000)(0.000000,-14.695706)
\psline[linewidth=\AxesLineWidth,linecolor=GridColor](0.500000,-15.000000)(0.500000,-14.695706)
\psline[linewidth=\AxesLineWidth,linecolor=GridColor](1.000000,-15.000000)(1.000000,-14.695706)
\psline[linewidth=\AxesLineWidth,linecolor=GridColor](1.500000,-15.000000)(1.500000,-14.695706)
\psline[linewidth=\AxesLineWidth,linecolor=GridColor](2.000000,-15.000000)(2.000000,-14.695706)
\psline[linewidth=\AxesLineWidth,linecolor=GridColor](0.000000,-15.000000)(0.024000,-15.000000)
\psline[linewidth=\AxesLineWidth,linecolor=GridColor](0.000000,-10.000000)(0.024000,-10.000000)
\psline[linewidth=\AxesLineWidth,linecolor=GridColor](0.000000,-5.000000)(0.024000,-5.000000)
\psline[linewidth=\AxesLineWidth,linecolor=GridColor](0.000000,0.000000)(0.024000,0.000000)
\psline[linewidth=\AxesLineWidth,linecolor=GridColor](0.000000,5.000000)(0.024000,5.000000)


{ \footnotesize 
\rput[t](0.000000,-15.304294){$0$}
\rput[t](0.500000,-15.304294){$0.5$}
\rput[t](1.000000,-15.304294){$1$}
\rput[t](1.500000,-15.304294){$1.5$}
\rput[t](2.000000,-15.304294){$2$}
\rput[r](-0.024000,-15.000000){$10^{-15}$}
\rput[r](-0.024000,-10.000000){$10^{-10}$}
\rput[r](-0.024000,-5.000000){$10^{-5}$}
\rput[r](-0.024000,0.000000){$10^{0}$}
\rput[r](-0.024000,5.000000){$10^{5}$}
} 

\psframe[linewidth=\AxesLineWidth,dimen=middle](0.000000,-15.000000)(2.000000,5.000000)

{ \small 
\rput[b](1.000000,-18){
\begin{tabular}{c}
$\tau$\\
\end{tabular}
}

\rput[t]{90}(-0.4,-5.000000){
\begin{tabular}{c}
relative frequency\\
\end{tabular}
}
} 

\newrgbcolor{color49.0046}{1  0  0}
\psline[plotstyle=line,linejoin=1,linestyle=solid,linewidth=\LineWidth,linecolor=color49.0046]
(0.020000,-1.694652)(0.060000,-0.751469)(0.100000,-0.329890)(0.140000,-0.070811)(0.180000,0.103241)
(0.220000,0.222245)(0.260000,0.300991)(0.300000,0.347872)(0.340000,0.368114)(0.380000,0.365190)
(0.420000,0.341529)(0.460000,0.298895)(0.500000,0.238609)(0.540000,0.161687)(0.580000,0.068926)
(0.620000,-0.039035)(0.660000,-0.161678)(0.700000,-0.298576)(0.740000,-0.449375)(0.780000,-0.613774)
(0.820000,-0.791520)(0.860000,-0.982395)(0.900000,-1.186209)(0.940000,-1.402801)(0.980000,-1.632025)
(1.020000,-1.873758)(1.060000,-2.127886)(1.100000,-2.394311)(1.140000,-2.672945)(1.180000,-2.963708)
(1.220000,-3.266529)(1.260000,-3.581343)(1.300000,-3.908093)(1.340000,-4.246724)(1.380000,-4.597189)
(1.420000,-4.959443)(1.460000,-5.333446)(1.500000,-5.719160)(1.540000,-6.116551)(1.580000,-6.525588)
(1.620000,-6.946241)(1.660000,-7.378483)(1.700000,-7.822288)(1.740000,-8.277634)(1.780000,-8.744498)
(1.820000,-9.222860)(1.860000,-9.712701)(1.900000,-10.214003)(1.940000,-10.726749)(1.980000,-11.250924)

\newrgbcolor{color50.0042}{0  0  1}
\psline[plotstyle=line,linejoin=1,showpoints=true,dotstyle=*,dotsize=\MarkerSize,linestyle=solid,linewidth=\LineWidth,linecolor=color50.0042]
(0.020000,-1.567271)(0.060000,-0.741908)(0.100000,-0.328288)(0.140000,-0.069106)(0.180000,0.104018)
(0.220000,0.221082)(0.260000,0.300076)(0.300000,0.347269)(0.340000,0.367109)(0.380000,0.364635)
(0.420000,0.340701)(0.460000,0.298500)(0.500000,0.238320)(0.540000,0.161107)(0.580000,0.069362)
(0.620000,-0.038196)(0.660000,-0.161960)(0.700000,-0.296569)(0.740000,-0.446824)(0.780000,-0.612744)
(0.820000,-0.789929)(0.860000,-0.979421)(0.900000,-1.177750)(0.940000,-1.403678)(0.980000,-1.621784)
(1.020000,-1.868141)(1.060000,-2.119901)(1.100000,-2.387746)(1.140000,-2.634980)(1.180000,-3.006564)
(1.220000,-3.288193)(1.260000,-3.530178)(1.300000,-4.000000)(1.340000,-4.170696)(1.380000,-4.602060)
(1.420000,-4.903090)(1.460000,-5.602060)(1.500000,-5.602060)

{ \small 
\rput[bl](0.048000,-14.391411){%
\psframebox[framesep=0pt,linewidth=\AxesLineWidth]{\psframebox*{\begin{tabular}{l}
\Rnode{a1}{\hspace*{0.0ex}} \hspace*{0.7cm} \Rnode{a2}{~~Maxwellian} \\
\Rnode{a3}{\hspace*{0.0ex}} \hspace*{0.7cm} \Rnode{a4}{~~Simulation} \\
\end{tabular}}
\ncline[linestyle=solid,linewidth=\LineWidth,linecolor=color49.0046]{a1}{a2}
\ncline[linestyle=solid,linewidth=\LineWidth,linecolor=color50.0042]{a3}{a4} \ncput{\psdot[dotstyle=*,dotsize=\MarkerSize,linecolor=color50.0042]}
}%
}%
} 

\end{pspicture}%
  \end{center}
  \caption{The PDF of the DGD of the second degree FIR filter compared against a true Maxwellian with the same mean.}
  \label{fig:cascade_dgd_pdf}
\end{figure}

Since $\mathbf{G}$ has real eigenvalues and orthogonal eigenvectors if it is constructed as in (\ref{eqn:structure_G}) and (\ref{eqn:components_G}) with $N=2$, it can be easily verified that $\mathbf{G}$ can be partitioned in the following way:

\begin{equation}
  \mathbf{G} = \mathbf{v}_1 \mathbf{v}_{1}^* + \mathbf{v}_2 \mathbf{v}_{2}^* , \qquad \mathbf{v}_{1}^* \mathbf{v}_1 = \mathbf{v}_{2}^* \mathbf{v}_2  = 1 \nonumber 
\end{equation}

\noindent with,
\begin{equation}
\begin{aligned}
  \mathbf{v}_1 &= \frac{1}{\sqrt{2}}\left( \sqrt{\lambda_1}\mathbf{e}_1 +  \sqrt{\lambda_2}\mathbf{e}_2 \right) \\
  \mathbf{v}_2 &= \frac{1}{\sqrt{2}}\left( \sqrt{\lambda_1}\mathbf{e}_1 -  \sqrt{\lambda_2}\mathbf{e}_2 \right)\quad.
\end{aligned}
  \label{eqn:eigVec_v_i}
\end{equation}

\noindent where $\mathbf{e}_i$ are the unit norm orthogonal eigenvectors of $\mathbf{G}$ with $\lambda_i$ as the corresponding eigenvalues.
In order to get a marginally uniform distribution for magnitude squares of the components of $\mathbf{v}_i$, the eigenvectors $\mathbf{e}_i$ can be multiplied with a phase term $e^{j\phi_i}$, where $\phi_i$ is selected uniformly at random in $[0,2\pi]$.
The resulting empirical DGD density obtained from a simulation with this selection method for $\mathbf{v}_i$ is displayed in Figure \ref{fig:cascade_dgd_pdf}. The sample size used in this simulation is $10^7$ and $\bar{\tau}$ was taken to be $0.4$.

Because of the positive definiteness constraint given in equation (\ref{eqn:sigma_constraints}), the above 2-section unit built with $\mathbf{v}_1$ and $\mathbf{v}_2$ can only reach DGD values up to a limit. This shortcoming of the second degree FIR filter can be overcome with the extension of the same idea to higher degree filters. 
The first possibility that comes to mind is to cascade M second degree filters and build a more general FIR filter with 2M sections.

\subsubsection{Extension to Higher Degree Filters}

\begin{figure*}
  \begin{center}
    \subfigure[PDF at $\omega = 0$]{
%
%
%
\providelength{\AxesLineWidth}       \setlength{\AxesLineWidth}{0.5pt}%
\providelength{\GridLineWidth}       \setlength{\GridLineWidth}{0.4pt}%
\providelength{\GridLineDotSep}      \setlength{\GridLineDotSep}{0.4pt}%
\providelength{\MinorGridLineWidth}  \setlength{\MinorGridLineWidth}{0.4pt}%
\providelength{\MinorGridLineDotSep} \setlength{\MinorGridLineDotSep}{0.8pt}%
\providelength{\plotwidth}           \setlength{\plotwidth}{6cm}
\providelength{\LineWidth}           \setlength{\LineWidth}{0.7pt}%
\providelength{\MarkerSize}          \setlength{\MarkerSize}{2pt}%
\newrgbcolor{GridColor}{0.8 0.8 0.8}%
%
\psset{xunit=0.126273\plotwidth,yunit=0.069715\plotwidth}%
\begin{pspicture}(-0.838710,-11.140351)(7.080645,0.263158)%


\psline[linestyle=dashed,dash=2pt 3pt,dotsep=\GridLineDotSep,linewidth=\GridLineWidth,linecolor=GridColor](0.000000,-10.000000)(0.000000,0.000000)
\psline[linestyle=dashed,dash=2pt 3pt,dotsep=\GridLineDotSep,linewidth=\GridLineWidth,linecolor=GridColor](1.000000,-10.000000)(1.000000,0.000000)
\psline[linestyle=dashed,dash=2pt 3pt,dotsep=\GridLineDotSep,linewidth=\GridLineWidth,linecolor=GridColor](2.000000,-10.000000)(2.000000,0.000000)
\psline[linestyle=dashed,dash=2pt 3pt,dotsep=\GridLineDotSep,linewidth=\GridLineWidth,linecolor=GridColor](3.000000,-10.000000)(3.000000,0.000000)
\psline[linestyle=dashed,dash=2pt 3pt,dotsep=\GridLineDotSep,linewidth=\GridLineWidth,linecolor=GridColor](4.000000,-10.000000)(4.000000,0.000000)
\psline[linestyle=dashed,dash=2pt 3pt,dotsep=\GridLineDotSep,linewidth=\GridLineWidth,linecolor=GridColor](5.000000,-10.000000)(5.000000,0.000000)
\psline[linestyle=dashed,dash=2pt 3pt,dotsep=\GridLineDotSep,linewidth=\GridLineWidth,linecolor=GridColor](6.000000,-10.000000)(6.000000,0.000000)
\psline[linestyle=dashed,dash=2pt 3pt,dotsep=\GridLineDotSep,linewidth=\GridLineWidth,linecolor=GridColor](7.000000,-10.000000)(7.000000,0.000000)
\psline[linestyle=dashed,dash=2pt 3pt,dotsep=\GridLineDotSep,linewidth=\GridLineWidth,linecolor=GridColor](0.000000,-10.000000)(7.000000,-10.000000)
\psline[linestyle=dashed,dash=2pt 3pt,dotsep=\GridLineDotSep,linewidth=\GridLineWidth,linecolor=GridColor](0.000000,-8.000000)(7.000000,-8.000000)
\psline[linestyle=dashed,dash=2pt 3pt,dotsep=\GridLineDotSep,linewidth=\GridLineWidth,linecolor=GridColor](0.000000,-6.000000)(7.000000,-6.000000)
\psline[linestyle=dashed,dash=2pt 3pt,dotsep=\GridLineDotSep,linewidth=\GridLineWidth,linecolor=GridColor](0.000000,-4.000000)(7.000000,-4.000000)
\psline[linestyle=dashed,dash=2pt 3pt,dotsep=\GridLineDotSep,linewidth=\GridLineWidth,linecolor=GridColor](0.000000,-2.000000)(7.000000,-2.000000)
\psline[linestyle=dashed,dash=2pt 3pt,dotsep=\GridLineDotSep,linewidth=\GridLineWidth,linecolor=GridColor](0.000000,0.000000)(7.000000,0.000000)


\psline[linewidth=\AxesLineWidth,linecolor=GridColor](0.000000,-10.000000)(0.000000,-9.847853)
\psline[linewidth=\AxesLineWidth,linecolor=GridColor](1.000000,-10.000000)(1.000000,-9.847853)
\psline[linewidth=\AxesLineWidth,linecolor=GridColor](2.000000,-10.000000)(2.000000,-9.847853)
\psline[linewidth=\AxesLineWidth,linecolor=GridColor](3.000000,-10.000000)(3.000000,-9.847853)
\psline[linewidth=\AxesLineWidth,linecolor=GridColor](4.000000,-10.000000)(4.000000,-9.847853)
\psline[linewidth=\AxesLineWidth,linecolor=GridColor](5.000000,-10.000000)(5.000000,-9.847853)
\psline[linewidth=\AxesLineWidth,linecolor=GridColor](6.000000,-10.000000)(6.000000,-9.847853)
\psline[linewidth=\AxesLineWidth,linecolor=GridColor](7.000000,-10.000000)(7.000000,-9.847853)
\psline[linewidth=\AxesLineWidth,linecolor=GridColor](0.000000,-10.000000)(0.084000,-10.000000)
\psline[linewidth=\AxesLineWidth,linecolor=GridColor](0.000000,-8.000000)(0.084000,-8.000000)
\psline[linewidth=\AxesLineWidth,linecolor=GridColor](0.000000,-6.000000)(0.084000,-6.000000)
\psline[linewidth=\AxesLineWidth,linecolor=GridColor](0.000000,-4.000000)(0.084000,-4.000000)
\psline[linewidth=\AxesLineWidth,linecolor=GridColor](0.000000,-2.000000)(0.084000,-2.000000)
\psline[linewidth=\AxesLineWidth,linecolor=GridColor](0.000000,0.000000)(0.084000,0.000000)


{ \footnotesize 
\rput[t](0.000000,-10.152147){$0$}
\rput[t](1.000000,-10.152147){$1$}
\rput[t](2.000000,-10.152147){$2$}
\rput[t](3.000000,-10.152147){$3$}
\rput[t](4.000000,-10.152147){$4$}
\rput[t](5.000000,-10.152147){$5$}
\rput[t](6.000000,-10.152147){$6$}
\rput[t](7.000000,-10.152147){$7$}
\rput[r](-0.084000,-10.000000){$10^{-10}$}
\rput[r](-0.084000,-8.000000){$10^{-8}$}
\rput[r](-0.084000,-6.000000){$10^{-6}$}
\rput[r](-0.084000,-4.000000){$10^{-4}$}
\rput[r](-0.084000,-2.000000){$10^{-2}$}
\rput[r](-0.084000,0.000000){$10^{0}$}
} 

\psframe[linewidth=\AxesLineWidth,dimen=middle](0.000000,-10.000000)(7.000000,0.000000)

{ \small 
\rput[b](3.500000,-11.8){
\begin{tabular}{c}
$\tau$\\
\end{tabular}
}

\rput[t]{90}(-1.7,-5.000000){
\begin{tabular}{c}
relative frequency\\
\end{tabular}
}
} 

\newrgbcolor{color505.0089}{1  0  0}
\psline[plotstyle=line,linejoin=1,linestyle=solid,linewidth=\LineWidth,linecolor=color505.0089]
(0.050000,-2.704110)(0.150000,-1.754187)(0.250000,-1.319130)(0.350000,-1.039834)(0.450000,-0.838825)
(0.550000,-0.686124)(0.650000,-0.566943)(0.750000,-0.472887)(0.850000,-0.398732)(0.950000,-0.341003)
(1.050000,-0.297272)(1.150000,-0.265775)(1.250000,-0.245190)(1.350000,-0.234503)(1.450000,-0.232914)
(1.550000,-0.239787)(1.650000,-0.254603)(1.750000,-0.276935)(1.850000,-0.306427)(1.950000,-0.342782)
(2.050000,-0.385743)(2.150000,-0.435094)(2.250000,-0.490646)(2.350000,-0.552236)(2.450000,-0.619720)
(2.550000,-0.692972)(2.650000,-0.771880)(2.750000,-0.856347)(2.850000,-0.946283)(2.950000,-1.041609)
(3.050000,-1.142253)(3.150000,-1.248152)(3.250000,-1.359247)(3.350000,-1.475484)(3.450000,-1.596816)
(3.550000,-1.723197)(3.650000,-1.854589)(3.750000,-1.990952)(3.850000,-2.132254)(3.950000,-2.278461)
(4.050000,-2.429545)(4.150000,-2.585480)(4.250000,-2.746238)(4.350000,-2.911798)(4.450000,-3.082137)
(4.550000,-3.257234)(4.650000,-3.437071)(4.750000,-3.621631)(4.850000,-3.810895)(4.950000,-4.004848)
(5.050000,-4.203476)(5.150000,-4.406765)(5.250000,-4.614701)(5.350000,-4.827272)(5.450000,-5.044467)
(5.550000,-5.266275)(5.650000,-5.492684)(5.750000,-5.723686)(5.850000,-5.959270)(5.950000,-6.199428)
(6.050000,-6.444152)(6.150000,-6.693433)(6.250000,-6.947264)(6.350000,-7.205637)(6.450000,-7.468545)
(6.550000,-7.735982)(6.650000,-8.007942)(6.750000,-8.284418)(6.850000,-8.565405)(6.950000,-8.850897)

\newrgbcolor{color506.0082}{0  0  1}
\psline[plotstyle=line,linejoin=1,showpoints=true,dotstyle=*,dotsize=\MarkerSize,linestyle=solid,linewidth=\LineWidth,linecolor=color506.0082]
(0.050000,-2.563837)(0.150000,-1.725380)(0.250000,-1.306097)(0.350000,-1.040005)(0.450000,-0.841426)
(0.550000,-0.684219)(0.650000,-0.568717)(0.750000,-0.476540)(0.850000,-0.400597)(0.950000,-0.341435)
(1.050000,-0.296038)(1.150000,-0.265136)(1.250000,-0.243623)(1.350000,-0.231488)(1.450000,-0.233067)
(1.550000,-0.241232)(1.650000,-0.258770)(1.750000,-0.279353)(1.850000,-0.303722)(1.950000,-0.343145)
(2.050000,-0.388266)(2.150000,-0.432997)(2.250000,-0.492914)(2.350000,-0.555190)(2.450000,-0.617029)
(2.550000,-0.689370)(2.650000,-0.771189)(2.750000,-0.850596)(2.850000,-0.951558)(2.950000,-1.047062)
(3.050000,-1.139482)(3.150000,-1.250109)(3.250000,-1.346595)(3.350000,-1.466355)(3.450000,-1.583526)
(3.550000,-1.720105)(3.650000,-1.850473)(3.750000,-1.982549)(3.850000,-2.142065)(3.950000,-2.319664)
(4.050000,-2.438899)(4.150000,-2.555955)(4.250000,-2.714443)(4.350000,-2.879426)(4.450000,-3.045757)
(4.550000,-3.221849)(4.650000,-3.337242)(4.750000,-3.769551)(4.850000,-3.886057)(4.950000,-4.000000)
(5.050000,-4.096910)(5.150000,-4.221849)(5.250000,-5.000000)(5.350000,-4.397940)(5.450000,-4.698970)
\psline[plotstyle=line,linejoin=1,showpoints=true,dotstyle=*,dotsize=\MarkerSize,linestyle=solid,linewidth=\LineWidth,linecolor=color506.0082]
(5.650000,-5.000000)(5.650000,-5.000000)
\psline[plotstyle=line,linejoin=1,showpoints=true,dotstyle=*,dotsize=\MarkerSize,linestyle=solid,linewidth=\LineWidth,linecolor=color506.0082]
(5.850000,-4.698970)(5.850000,-4.698970)

{ \small 
\rput[bl](0.168000,-9.695706){%
\psframebox[framesep=0pt,linewidth=\AxesLineWidth]{\psframebox*{\begin{tabular}{l}
\Rnode{a1}{\hspace*{0.0ex}} \hspace*{0.7cm} \Rnode{a2}{~~Maxwellian} \\
\Rnode{a3}{\hspace*{0.0ex}} \hspace*{0.7cm} \Rnode{a4}{~~Simulation} \\
\end{tabular}}
\ncline[linestyle=solid,linewidth=\LineWidth,linecolor=color505.0089]{a1}{a2}
\ncline[linestyle=solid,linewidth=\LineWidth,linecolor=color506.0082]{a3}{a4} \ncput{\psdot[dotstyle=*,dotsize=\MarkerSize,linecolor=color506.0082]}
}%
}%
} 

\end{pspicture}%
       \label{fig:cascade_dgd_pdf_0}
     }
     \subfigure[PDF at $\omega = \frac{\pi}{4}$]{
%
%
%
\providelength{\AxesLineWidth}       \setlength{\AxesLineWidth}{0.5pt}%
\providelength{\GridLineWidth}       \setlength{\GridLineWidth}{0.4pt}%
\providelength{\GridLineDotSep}      \setlength{\GridLineDotSep}{0.4pt}%
\providelength{\MinorGridLineWidth}  \setlength{\MinorGridLineWidth}{0.4pt}%
\providelength{\MinorGridLineDotSep} \setlength{\MinorGridLineDotSep}{0.8pt}%
\providelength{\plotwidth}           \setlength{\plotwidth}{6cm}
\providelength{\LineWidth}           \setlength{\LineWidth}{0.7pt}%
\providelength{\MarkerSize}          \setlength{\MarkerSize}{2pt}%
\newrgbcolor{GridColor}{0.8 0.8 0.8}%
%
\psset{xunit=0.126273\plotwidth,yunit=0.069715\plotwidth}%
\begin{pspicture}(-0.838710,-11.140351)(7.080645,0.263158)%


\psline[linestyle=dashed,dash=2pt 3pt,dotsep=\GridLineDotSep,linewidth=\GridLineWidth,linecolor=GridColor](0.000000,-10.000000)(0.000000,0.000000)
\psline[linestyle=dashed,dash=2pt 3pt,dotsep=\GridLineDotSep,linewidth=\GridLineWidth,linecolor=GridColor](1.000000,-10.000000)(1.000000,0.000000)
\psline[linestyle=dashed,dash=2pt 3pt,dotsep=\GridLineDotSep,linewidth=\GridLineWidth,linecolor=GridColor](2.000000,-10.000000)(2.000000,0.000000)
\psline[linestyle=dashed,dash=2pt 3pt,dotsep=\GridLineDotSep,linewidth=\GridLineWidth,linecolor=GridColor](3.000000,-10.000000)(3.000000,0.000000)
\psline[linestyle=dashed,dash=2pt 3pt,dotsep=\GridLineDotSep,linewidth=\GridLineWidth,linecolor=GridColor](4.000000,-10.000000)(4.000000,0.000000)
\psline[linestyle=dashed,dash=2pt 3pt,dotsep=\GridLineDotSep,linewidth=\GridLineWidth,linecolor=GridColor](5.000000,-10.000000)(5.000000,0.000000)
\psline[linestyle=dashed,dash=2pt 3pt,dotsep=\GridLineDotSep,linewidth=\GridLineWidth,linecolor=GridColor](6.000000,-10.000000)(6.000000,0.000000)
\psline[linestyle=dashed,dash=2pt 3pt,dotsep=\GridLineDotSep,linewidth=\GridLineWidth,linecolor=GridColor](7.000000,-10.000000)(7.000000,0.000000)
\psline[linestyle=dashed,dash=2pt 3pt,dotsep=\GridLineDotSep,linewidth=\GridLineWidth,linecolor=GridColor](0.000000,-10.000000)(7.000000,-10.000000)
\psline[linestyle=dashed,dash=2pt 3pt,dotsep=\GridLineDotSep,linewidth=\GridLineWidth,linecolor=GridColor](0.000000,-8.000000)(7.000000,-8.000000)
\psline[linestyle=dashed,dash=2pt 3pt,dotsep=\GridLineDotSep,linewidth=\GridLineWidth,linecolor=GridColor](0.000000,-6.000000)(7.000000,-6.000000)
\psline[linestyle=dashed,dash=2pt 3pt,dotsep=\GridLineDotSep,linewidth=\GridLineWidth,linecolor=GridColor](0.000000,-4.000000)(7.000000,-4.000000)
\psline[linestyle=dashed,dash=2pt 3pt,dotsep=\GridLineDotSep,linewidth=\GridLineWidth,linecolor=GridColor](0.000000,-2.000000)(7.000000,-2.000000)
\psline[linestyle=dashed,dash=2pt 3pt,dotsep=\GridLineDotSep,linewidth=\GridLineWidth,linecolor=GridColor](0.000000,0.000000)(7.000000,0.000000)


\psline[linewidth=\AxesLineWidth,linecolor=GridColor](0.000000,-10.000000)(0.000000,-9.847853)
\psline[linewidth=\AxesLineWidth,linecolor=GridColor](1.000000,-10.000000)(1.000000,-9.847853)
\psline[linewidth=\AxesLineWidth,linecolor=GridColor](2.000000,-10.000000)(2.000000,-9.847853)
\psline[linewidth=\AxesLineWidth,linecolor=GridColor](3.000000,-10.000000)(3.000000,-9.847853)
\psline[linewidth=\AxesLineWidth,linecolor=GridColor](4.000000,-10.000000)(4.000000,-9.847853)
\psline[linewidth=\AxesLineWidth,linecolor=GridColor](5.000000,-10.000000)(5.000000,-9.847853)
\psline[linewidth=\AxesLineWidth,linecolor=GridColor](6.000000,-10.000000)(6.000000,-9.847853)
\psline[linewidth=\AxesLineWidth,linecolor=GridColor](7.000000,-10.000000)(7.000000,-9.847853)
\psline[linewidth=\AxesLineWidth,linecolor=GridColor](0.000000,-10.000000)(0.084000,-10.000000)
\psline[linewidth=\AxesLineWidth,linecolor=GridColor](0.000000,-8.000000)(0.084000,-8.000000)
\psline[linewidth=\AxesLineWidth,linecolor=GridColor](0.000000,-6.000000)(0.084000,-6.000000)
\psline[linewidth=\AxesLineWidth,linecolor=GridColor](0.000000,-4.000000)(0.084000,-4.000000)
\psline[linewidth=\AxesLineWidth,linecolor=GridColor](0.000000,-2.000000)(0.084000,-2.000000)
\psline[linewidth=\AxesLineWidth,linecolor=GridColor](0.000000,0.000000)(0.084000,0.000000)


{ \footnotesize 
\rput[t](0.000000,-10.152147){$0$}
\rput[t](1.000000,-10.152147){$1$}
\rput[t](2.000000,-10.152147){$2$}
\rput[t](3.000000,-10.152147){$3$}
\rput[t](4.000000,-10.152147){$4$}
\rput[t](5.000000,-10.152147){$5$}
\rput[t](6.000000,-10.152147){$6$}
\rput[t](7.000000,-10.152147){$7$}
\rput[r](-0.084000,-10.000000){$10^{-10}$}
\rput[r](-0.084000,-8.000000){$10^{-8}$}
\rput[r](-0.084000,-6.000000){$10^{-6}$}
\rput[r](-0.084000,-4.000000){$10^{-4}$}
\rput[r](-0.084000,-2.000000){$10^{-2}$}
\rput[r](-0.084000,0.000000){$10^{0}$}
} 

\psframe[linewidth=\AxesLineWidth,dimen=middle](0.000000,-10.000000)(7.000000,0.000000)

{ \small 
\rput[b](3.500000,-11.8){
\begin{tabular}{c}
$\tau$\\
\end{tabular}
}

\rput[t]{90}(-1.7,-5.000000){
\begin{tabular}{c}
relative frequency\\
\end{tabular}
}
} 

\newrgbcolor{color501.0088}{1  0  0}
\psline[plotstyle=line,linejoin=1,linestyle=solid,linewidth=\LineWidth,linecolor=color501.0088]
(0.050000,-2.704110)(0.150000,-1.754187)(0.250000,-1.319130)(0.350000,-1.039834)(0.450000,-0.838825)
(0.550000,-0.686124)(0.650000,-0.566943)(0.750000,-0.472887)(0.850000,-0.398732)(0.950000,-0.341003)
(1.050000,-0.297272)(1.150000,-0.265775)(1.250000,-0.245190)(1.350000,-0.234503)(1.450000,-0.232914)
(1.550000,-0.239787)(1.650000,-0.254603)(1.750000,-0.276935)(1.850000,-0.306427)(1.950000,-0.342782)
(2.050000,-0.385743)(2.150000,-0.435094)(2.250000,-0.490646)(2.350000,-0.552236)(2.450000,-0.619720)
(2.550000,-0.692972)(2.650000,-0.771880)(2.750000,-0.856347)(2.850000,-0.946283)(2.950000,-1.041609)
(3.050000,-1.142253)(3.150000,-1.248152)(3.250000,-1.359247)(3.350000,-1.475484)(3.450000,-1.596816)
(3.550000,-1.723197)(3.650000,-1.854589)(3.750000,-1.990952)(3.850000,-2.132254)(3.950000,-2.278461)
(4.050000,-2.429545)(4.150000,-2.585480)(4.250000,-2.746238)(4.350000,-2.911798)(4.450000,-3.082137)
(4.550000,-3.257234)(4.650000,-3.437071)(4.750000,-3.621631)(4.850000,-3.810895)(4.950000,-4.004848)
(5.050000,-4.203476)(5.150000,-4.406765)(5.250000,-4.614701)(5.350000,-4.827272)(5.450000,-5.044467)
(5.550000,-5.266275)(5.650000,-5.492684)(5.750000,-5.723686)(5.850000,-5.959270)(5.950000,-6.199428)
(6.050000,-6.444152)(6.150000,-6.693433)(6.250000,-6.947264)(6.350000,-7.205637)(6.450000,-7.468545)
(6.550000,-7.735982)(6.650000,-8.007942)(6.750000,-8.284418)(6.850000,-8.565405)(6.950000,-8.850897)

\newrgbcolor{color502.0083}{0  0  1}
\psline[plotstyle=line,linejoin=1,showpoints=true,dotstyle=*,dotsize=\MarkerSize,linestyle=solid,linewidth=\LineWidth,linecolor=color502.0083]
(0.050000,-2.565431)(0.150000,-1.745694)(0.250000,-1.322211)(0.350000,-1.041484)(0.450000,-0.841155)
(0.550000,-0.683673)(0.650000,-0.568459)(0.750000,-0.473803)(0.850000,-0.397051)(0.950000,-0.339998)
(1.050000,-0.298164)(1.150000,-0.267405)(1.250000,-0.247406)(1.350000,-0.234562)(1.450000,-0.234175)
(1.550000,-0.239864)(1.650000,-0.252643)(1.750000,-0.280810)(1.850000,-0.308618)(1.950000,-0.340654)
(2.050000,-0.384808)(2.150000,-0.433904)(2.250000,-0.489589)(2.350000,-0.549397)(2.450000,-0.619355)
(2.550000,-0.692697)(2.650000,-0.772190)(2.750000,-0.854524)(2.850000,-0.945693)(2.950000,-1.038484)
(3.050000,-1.134541)(3.150000,-1.244888)(3.250000,-1.357041)(3.350000,-1.477686)(3.450000,-1.595166)
(3.550000,-1.719422)(3.650000,-1.871601)(3.750000,-1.969805)(3.850000,-2.106238)(3.950000,-2.302771)
(4.050000,-2.408935)(4.150000,-2.600326)(4.250000,-2.752027)(4.350000,-2.847712)(4.450000,-3.193820)
(4.550000,-3.251812)(4.650000,-3.443697)(4.750000,-3.537602)(4.850000,-3.744727)(4.950000,-3.886057)
(5.050000,-4.301030)(5.150000,-4.301030)(5.250000,-4.698970)(5.350000,-4.522879)(5.450000,-4.698970)
\psline[plotstyle=line,linejoin=1,showpoints=true,dotstyle=*,dotsize=\MarkerSize,linestyle=solid,linewidth=\LineWidth,linecolor=color502.0083]
(5.650000,-4.698970)(5.750000,-5.000000)

{ \small 
\rput[bl](0.168000,-9.695706){%
\psframebox[framesep=0pt,linewidth=\AxesLineWidth]{\psframebox*{\begin{tabular}{l}
\Rnode{a1}{\hspace*{0.0ex}} \hspace*{0.7cm} \Rnode{a2}{~~Maxwellian} \\
\Rnode{a3}{\hspace*{0.0ex}} \hspace*{0.7cm} \Rnode{a4}{~~Simulation} \\
\end{tabular}}
\ncline[linestyle=solid,linewidth=\LineWidth,linecolor=color501.0088]{a1}{a2}
\ncline[linestyle=solid,linewidth=\LineWidth,linecolor=color502.0083]{a3}{a4} \ncput{\psdot[dotstyle=*,dotsize=\MarkerSize,linecolor=color502.0083]}
}%
}%
} 

\end{pspicture}%
       \label{fig:cascade_dgd_pdf_pib4}
     }
  \end{center}
  \caption{PDF of the DGD at the center frequency ($\omega = 0$) and the corner frequency ($\omega = \frac{\pi}{4}$) for the CSM. Solid curves represent the Maxwellian with mean 1.6.}
  \label{fig:cascade_dgd_pdfs}
\end{figure*}
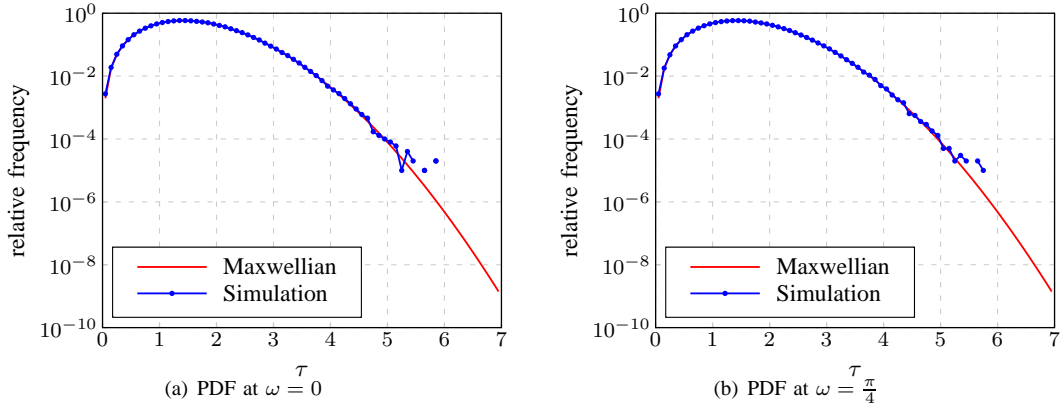

Using the fact that the sum of two independent Gaussian random variables is again a Gaussian random variable, we can extend the second degree FIR filter described above to a more general system of even degree.

\begin{equation}
  \mathbf{G} = \mathbf{G}_1+\mathbf{G}_2+ \dots +\mathbf{G}_M \nonumber \nonumber
\end{equation}
$\mathbf{G}$ has a Maxwellian spacing distribution if the summands $\mathbf{G}_i$ are constructed independently according to (\ref{eqn:structure_G}) and (\ref{eqn:components_G}). 
In this case, the unit norm vectors $\mathbf{v}_i$ can be selected with the same scheme as (\ref{eqn:eigVec_v_i}):

\begin{equation}
  \mathbf{G} = \sum_{k=1}^M \left[
 	\begin{array}[pos]{cc}
		A_k & C_k \\
      		C_k^* & B_k 
	\end{array} 
	\right] = \sum_{k_1,k_2=1}^M \mathbf{v}_{k_1} \mathbf{v}_{k_1}^* + \mathbf{v}_{k_2} \mathbf{v}_{k_2}^*
    \label{eqn:structure_sum_G}
\end{equation}

\begin{equation}
	\begin{array}{cc}
	  A_k \sim \mathcal{N}(1,\sigma) & \Re{\{C_k\}} \sim  \mathcal{N}(0,\sigma) \\
	  B_k = N-a_k & \Im{\{C_k\}} \sim \mathcal{N}(0, \sigma)   
	\end{array}
  \label{eqn:components_sum_G}
\end{equation}

\noindent This filter of degree 2M will have the following mean DGD:

\begin{equation}
  \bar{\tau} = 2\sqrt{\frac{8M}{\pi}}\sigma \quad.
  \label{eqn:tau_cascade}
\end{equation}

\subsubsection{Constraints on Model Parameters}
\label{DGD_constraints}

As discussed earlier, the selection of the standard deviation $\sigma$ in (\ref{eqn:components_G}) is not arbitrary. 
The fact that $\mathbf{G}$ must be positive  definite restrains $\sigma$ with an upper bound determined by the minimum probability value we want to match in the DGD distribution.
For a fixed value of $\sigma$ the probability that $\mathbf{G}$ in (\ref{eqn:structure_G}) is not positive definite can be computed from (\ref{eqn:sigma_constraints}).

The last constraint in (\ref{eqn:sigma_constraints}) takes precedence over the other two since $A$ and $B$ can never be  negative simultaneously and $|C|^2$ is always positive. 
The maximum value $AB = A(N-A)$ can take is $\frac{N^2}{4}$.  
Hence in the case that $|C| \geq \frac{N}{2}$, $\mathbf{G}$ is never positive definite. 
Similarly for $A \notin (0,N)$ $\mathbf{G}$ cannot be positive definite.
This restricts the range of $\tau$ to $(0, N)$ as it is expected from an FIR filter built with N delay elements.

With the conditions above we can calculate the probability that $\mathbf{G}$ is  positive definite, $\mathrm{P}(\mathbf{G} \succ 0)$, with given $N$ and $\sigma$, 

\begin{equation}
   P(\{A = a \, | \, a \in (0,N)\} \cap \{|C|=c\,|\,c^2 \leq a(N-a)\})  \;.
\end{equation}
After a bit of fiddling we arrive at,

\begin{equation}
  \mathrm{P}(\mathbf{G} \succ 0) = \mathrm{erf}\left(\frac{N}{2\sqrt{2}\sigma}\right) - \frac{N}{\sqrt{2\pi}\sigma} e^{\frac{-N^2}{8\sigma^2}} \;. \nonumber
\end{equation}

\noindent where $\mathrm{erf}(\bullet)$ is the error function.
Note that this expression is equal to the probability that a Maxwellian  random variable with the same mean as in (\ref{eqn:mean_tau_of_sigma}) is smaller than $N$. In other words, the positive definiteness of $\mathbf{G}$ ensures that the resulting DGD is at most $N$ which is the maximum delay of a discrete time filter with $N$ sections.
This result can be used as a filter degree selection tool for a given minimum probability we want to match, $p$, and a mean DGD, $\bar{\tau}$.

\begin{eqnarray}
  p &=& 1 - \mathrm{P}(\mathbf{G} \succ 0) \nonumber \\
  &=& \mathrm{erfc}\left(\frac{N}{2\sqrt{2}\sigma}\right) + \frac{N}{\sqrt{2\pi}\sigma} e^{\frac{-N^2}{8\sigma^2}} \nonumber \\
  &=& 2\left[\mathrm{Q}\left(\frac{N}{2\sigma}\right) + \frac{N}{\sqrt{2\pi}2\sigma}e^{-\frac{(N/2\sigma)^2}{2}}\right]
  \label{eqn:prob_p}
\end{eqnarray}
where $Q(\bullet)$ is the Q-function of the standard normal distribution.

\subsubsection{Frequency Behavior of the Cascading Method}

The analysis given above only discusses the properties of the cascaded sampling method at the center frequency but the reason this method was chosen among other alternatives to partition a positive definite matrix as sum of idempotents is its uniform frequency behavior.

\begin{figure}[t]
  \begin{center}
    \input{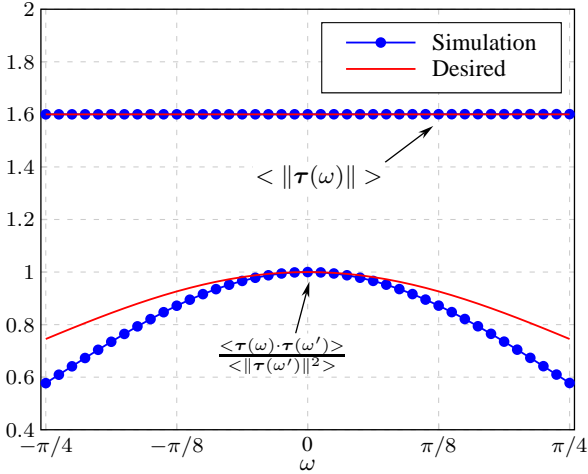}
  \end{center}
  \caption{Mean and normalized autocorrelation curves of the cascaded sampling method compared with the expected values.}
  \label{fig:cascade_mean_corr}
\end{figure}

Figure \ref{fig:cascade_dgd_pdfs} illustrates this behavior in terms of the PDF of the DGD. These graphics display the DGD distribution of a filter with 20 sections and a mean DGD value of 1.6 first at the center frequency, $\omega = 0$, and then at the corner frequency, $\omega = \frac{\pi}{4}$. The solid curves in the graphics are the expected Maxwellians with mean 1.6. Figure \ref{fig:cascade_dgd_pdfs} shows complete agreement of the model output with the desired values at the center as well as the corner frequency. This behavior is typical for the CSM with low enough mean DGD values, however, frequency dependent statistics do not display a similar behavior. Figure \ref{fig:cascade_mean_corr} shows that although the mean of the DGD distribution is constant over the whole frequency range, the correlation structure deviates from the desired curve. Therefore the CSM, in its current form, remains only as a tool for accurately matching the first-order behavior of a real PMD channel.

\subsection{Compensated MCMC Method}

MCMC algorithms are based on the construction of a Markov chain on the sample space  of a general vector random variable $\mathbf{X}$. 
In typical settings, the probability density function, $f_{\mathbf{X}}(\mathbf{x})$, of the distribution can only be evaluated up to a normalizing constant. 
The common Metropolis algorithm \cite{metropolis1953equation} starts with an initial state, $\mathbf{x}$, and generates samples of the random variable iteratively. 
At every step of the procedure, a new state, $\mathbf{x}'$, is proposed according to some proposal distribution $p(\mathbf{x},\mathbf{x}')$. 
This proposal state is then accepted with a probability determined by the ratio of the PDF values for the new state and the old state.

 \begin{equation}
   \alpha(\mathbf{x},\mathbf{x}') = \mathrm{min}\left(1, \frac{f_{\mathbf{X}}(\mathbf{x}')}{f_{\mathbf{X}}(\mathbf{x})} \right)
   \label{eqn:accpt-rejct}
 \end{equation}

Because the accept-reject rule only requires the evaluation of the ratio of the probability densities for the proposed and the old state, it is sufficient to know the target PDF up to a scalar constant. 
The sole restriction of the  Metropolis algorithm is that the proposal density be symmetric and simple enough to sample directly.

One generalization of the Metropolis algorithm is the Metropolis-Hastings algorithm \cite{hastings1970monte} which can employ asymmetric proposal densities. 
In order to achieve this, the acceptance probability is modified so as to incorporate a ratio of the proposal density values.

 \begin{equation}
   \alpha(\mathbf{x},\mathbf{x}') = \mathrm{min}\left(1, \frac{f_{\mathbf{X}}(\mathbf{x}')}{f_{\mathbf{X}}(\mathbf{x})} \frac{p(\mathbf{x}',\mathbf{x})}{p(\mathbf{x},\mathbf{x}')}\right)
   \label{eqn:mh-accpt-rejct}
 \end{equation}

\begin{figure*}
  \begin{center}
    \subfigure[Joint distribution of the first and second components at $f=0$]{
    \includegraphics[width=0.45\textwidth]{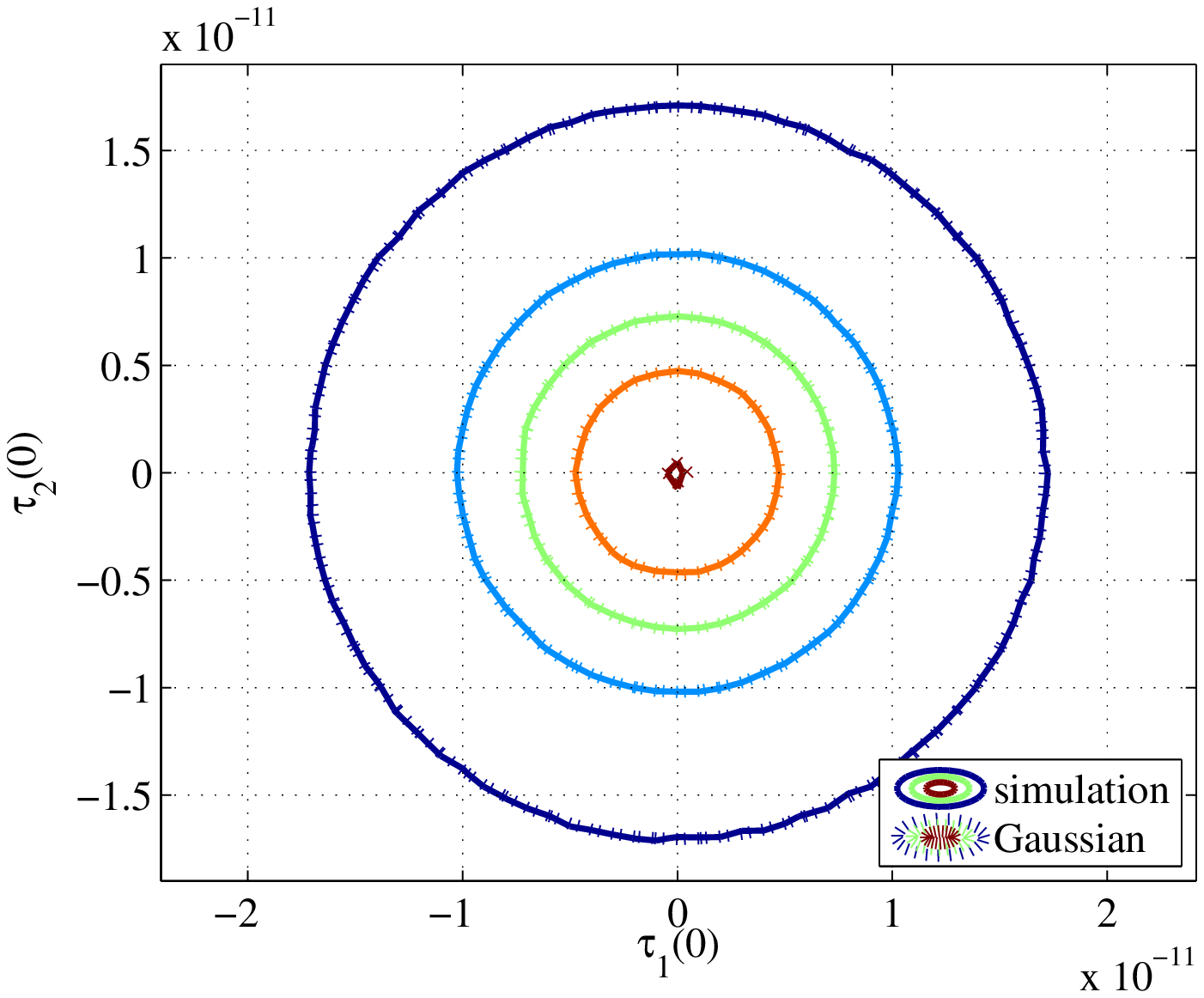}
    \label{fig:jointCont2}}
    \subfigure[Joint distribution of the first components at $f=0$  and $f=20$ GHz]{
    \includegraphics[width=0.45\textwidth]{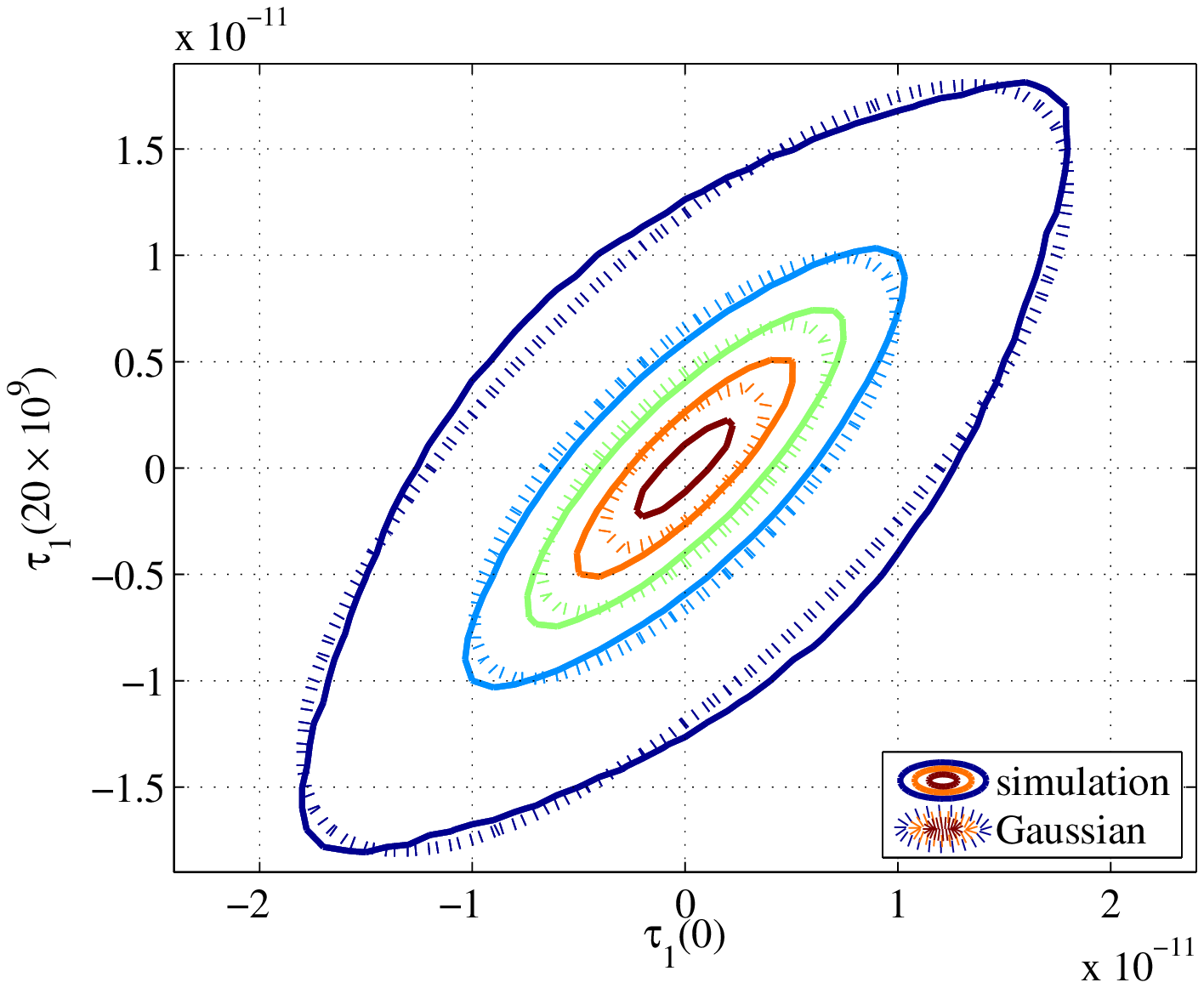}
    \label{fig:jointCont}}
  \end{center}
  \caption{Contour plots of the joint distributions of PMD vector components}
  \label{fig:jointConts}
\end{figure*}

Now let us consider how MCMC can be employed to generate samples of fiber models using paraunitary FIR filters. 
A discrete time fiber model with $N$ concatenated degree-one sections can be viewed as a complex mapping from the sample space of filter parameters to the space of PMD vector values. 
This mapping accepts a set of $2 \times 1$ complex valued unit norm vectors,  $\{\mathbf{v}_1, \mathbf{v}_2, \dots, \mathbf{v}_N\}$, which have a total of $2N$ real scalar parameters as input and produces a frequency dependent PMD vector, $\boldsymbol{\tau}(\omega)$, at the output. 
If we discretize the frequency axis such that we force the statistical properties of the PMD vectors, $\{\boldsymbol{\tau}(\omega_1), \boldsymbol{\tau}(\omega_2), \dots \boldsymbol{\tau}(\omega_M)\}$ at a set of frequencies, $\omega_1$ through $\omega_M$, in the frequency range of interest, we can expect the model to behave similarly at intermediate frequencies. 
Consequently, we obtain a mapping from $\mathbb{R}^{2N}$ to $\mathbb{R}^M$. Therefore, the problem of sampling the input parameters, such that the output statistics exhibit the desired behavior, can be described as follows.

Suppose we are given a general many-to-one, non-isometric map $h: \mathbb{R}^n \to \mathbb{R}^m$ which maps a random vector $\mathbf{X}$ to another random vector $\mathbf{Y} = h(\mathbf{X})$ where $\mathbf{X} \in \mathbb{R}^n$, $\mathbf{Y} \in \mathbb{R}^m$ and let $f_{\mathbf{Y}d}$ be the desired probability density of $\mathbf{Y}$. 
Given $f_{\mathbf{Y}d}$ how must $f_{\mathbf{X}}$ be chosen such that the transformed variable $\mathbf{Y} = h(\mathbf{X})$ has the desired PDF?

The answer to this question was given in \cite{mahmutoglu_random_2012} under the framework of the compensated MCMC algorithm which modifies the accept-reject rule in the standard Metropolis-Hastings algorithm as

\begin{equation}
  \alpha_c(\mathbf{x},\mathbf{x}') = \mathrm{min}\left(1, \frac{f_{\mathbf{Y}d}(h(\mathbf{x}'))}{f_{\mathbf{Y}d}(h(\mathbf{x}))} \frac{p(\mathbf{x}',\mathbf{x})}{p(\mathbf{x},\mathbf{x}')}
  \frac{f_{U}(h(\mathbf{x}))}{f_{U}(h(\mathbf{x}'))} \right)
  \label{eq:mod-accpt-rejct} 
\end{equation}

\noindent where $f_U(\bullet)$ is the distribution of a set of output random vectors consisting of resulting PMD vectors at chosen frequencies when the model parameters are sampled uniformly and independently. This distribution will be called uniform parameter distribution and abbreviated as UPD in the following.

Unlike the cascaded sampling method, where we tried to match only the DGD distribution, with the C-MCMC method we strive to match the frequency dependent statistics of a true PMD channel. In order to achieve this goal, we use the output PMD vector of the model as the target random vector. 
Due to the complicated nature of the PMD vector, we make use of simplifying assumptions and approximations about its probability distributions concerning the full model as well as the paraunitary FIR filter.

\subsubsection{PMD Vector and its Joint PDF at a Particular Frequency}
In  the literature, polarization mode dispersion is described in terms of the PMD vector $\boldsymbol{\tau}(\omega) = \tau(\omega)\mathbf{p}(\omega)$. 
Here $\tau$ is the differential group delay between the fast and slow polarization directions and $\mathbf{p}$ is the unit vector pointing in the direction of the slow PSP.
The PMD vector is defined as the differential change a polarization state at the output of the fiber undergoes as the angular frequency varies. 
\[
\frac{\mathrm{d}\mathbf{s}}{\mathrm{d}\omega} = \boldsymbol{\tau} \times \boldsymbol{s} \quad,
\]

\noindent where $\mathbf{s}$ is a polarization state at the output.

It can be shown theoretically as well as experimentally that $\boldsymbol{\tau}$ is the result of a random walk process in a three dimensional space and hence has i.i.d. zero-mean Gaussian vector components \cite{damask2005polarization}. 
As a result of this statistical property, the Euclidean norm of the PMD vector $\tau$ has a Maxwellian distribution.

\subsubsection{Frequency Dependent Behavior of the PMD Vector}

A fiber, subject to polarization mode dispersion, can be modeled as the concatenation of many statistically independent birefringent sections. As the number of sections grow, the accuracy of the model increases. 
In order to investigate the joint statistical properties of PMD vectors at different frequencies, a fiber was simulated using 200 sections and $4.5 \times 10^6$ samples. The mean DGD was set to be 10 psec. 

There is no precise theoretical result in the literature about the joint distribution of the components of the PMD vector $\boldsymbol{\tau}$ at different frequencies. It is known that the marginals of these components are Gaussian and independent at the same frequency. Figure \ref{fig:jointCont2} shows the contours of the joint distribution of the first and second components of $\boldsymbol{\tau}$. The dotted lines are the contours of a jointly Gaussian distribution fitted to the simulation data. In this case, the covariance matrix is
\[
C = \left[ \begin{array}{cc}
  0.3943 & 0.0001 \\
  0.0001 & 0.3921 \end{array}\right] \times 10^{-22}
\]

\noindent which is in agreement with the expected result.

On the other hand, one expects the i$^{\text{th}}$ component of $\boldsymbol{\tau}$ to be dependent on the i$^{\text{th}}$ component at another frequency. 
The joint distribution is again unknown but there is evidence suggesting that it is not jointly Gaussian due to its frequency derivative not being Gaussian \cite{foschini1991statistical}. 
This odd behavior of random variables with Gaussian marginals not being jointly Gaussian can be seen in Figure \ref{fig:jointCont}. 
Although there is a clear deviation from the fitted jointly Gaussian distribution, the similarity between  two curves hints to an approximation with a jointly Gaussian distribution. The covariance matrix computed for the fitted Gaussian this time is

\[
C = \left[ \begin{array}{cc}
    0.3943 &  0.2951 \\
     0.2951 & 0.3939 
  \end{array}\right] \times 10^{-22} \quad .
\]

\begin{figure*}
  \begin{center}
    \begin{pspicture}[showgrid=false](-8,0)(8,5)
  \psset{xunit=0.9cm,yunit=0.5cm}
    \rput(0,0){\pnode(0,0){1b}
      \rput(0,1){\ovalnode{1c11}{$c_2$}}
      \pnode(-1,1){1l1}
      \pnode(1,1){1r1}
      \rput(-1,2){\ovalnode{1c21}{$c_1$}}
      \rput(1,2){\ovalnode{1c22}{$c_1$}}
      \pnode(-2,2){1l2}
      \pnode(2,2){1r2}
      \pnode(-3,3){1l3}
      \pnode(3,3){1r3}
      \rput(-2,3){\ovalnode{1c31}{$I$}}
      \rput(0,3){\ovalnode{1c32}{$I$}}
      \rput(2,3){\ovalnode{1c33}{$I$}}
      
      \pnode(0,6){1t}
      \rput(0,5){\ovalnode{1c51}{$c_2$}}
      \pnode(-1,5){1l5}
      \pnode(1,5){1r5}
      \rput(-1,4){\ovalnode{1c41}{$c_1$}}
      \rput(1,4){\ovalnode{1c42}{$c_1$}}
      \pnode(-2,4){1l4}
      \pnode(2,4){1r4}
    }

  \ncline{1b}{1l1}
  \ncline{1l1}{1l2}
  \ncline{1b}{1r1}
  \ncline{1r1}{1r2}
  \ncline{1r2}{1r3}
  \ncline{1l2}{1l3}
  \ncline{1c21}{1c22}
  \ncline{1c31}{1c32}
  \ncline{1c32}{1c33}
  
  \ncline{1t}{1l5}
  \ncline{1l5}{1l4}
  \ncline{1t}{1r5}
  \ncline{1r5}{1r4}
  \ncline{1r4}{1r3}
  \ncline{1l4}{1l3}
  \ncline{1c41}{1c42} 
  
  \rput(-3,3){\pnode(0,0){2b}
      \rput(0,1){\ovalnode{2c11}{$c_2$}}
      \pnode(-1,1){2l1}
      \pnode(1,1){2r1}
      \rput(-1,2){\ovalnode{2c21}{$c_1$}}
      \rput(1,2){\ovalnode{2c22}{$c_1$}}
      \pnode(-2,2){2l2}
      \pnode(2,2){2r2}
      \pnode(-3,3){2l3}
      \pnode(3,3){2r3}
      \rput(-2,3){\ovalnode{2c31}{$I$}}
      \rput(0,3){\ovalnode{2c32}{$I$}}
      \rput(2,3){\ovalnode{2c33}{$I$}}
      
      \pnode(0,6){2t}
      \rput(0,5){\ovalnode{2c51}{$c_2$}}
      \pnode(-1,5){2l5}
      \pnode(1,5){2r5}
      \rput(-1,4){\ovalnode{2c41}{$c_1$}}
      \rput(1,4){\ovalnode{2c42}{$c_1$}}
      \pnode(-2,4){2l4}
      \pnode(2,4){2r4}
    }

  \ncline{2b}{2l1}
  \ncline{2l1}{2l2}
  \ncline{2l2}{2l3}
  \ncline{2c21}{2c22}
  \ncline{2c31}{2c32}
  \ncline{2c32}{2c33}
  
  \ncline{2t}{2l5}
  \ncline{2l5}{2l4}
  \ncline{2t}{2r5}
  \ncline{2r5}{2r4}
  \ncline{2r4}{2r3}
  \ncline{2l4}{2l3}
  \ncline{2c41}{2c42}

  \rput(3,3){\pnode(0,0){3b}
      \rput(0,1){\ovalnode{3c11}{$c_2$}}
      \pnode(-1,1){3l1}
      \pnode(1,1){3r1}
      \rput(-1,2){\ovalnode{3c21}{$c_1$}}
      \rput(1,2){\ovalnode{3c22}{$c_1$}}
      \pnode(-2,2){3l2}
      \pnode(2,2){3r2}
      \pnode(-3,3){3l3}
      \pnode(3,3){3r3}
      \rput(-2,3){\ovalnode{3c31}{$I$}}
      \rput(0,3){\ovalnode{3c32}{$I$}}
      \rput(2,3){\ovalnode{3c33}{$I$}}
      
      \pnode(0,6){3t}
      \rput(0,5){\ovalnode{3c51}{$c_2$}}
      \pnode(-1,5){3l5}
      \pnode(1,5){3r5}
      \rput(-1,4){\ovalnode{3c41}{$c_1$}}
      \rput(1,4){\ovalnode{3c42}{$c_1$}}
      \pnode(-2,4){3l4}
      \pnode(2,4){3r4}
    }

  \ncline{3b}{3r1}
  \ncline{3r1}{3r2}
  \ncline{3r2}{3r3}
  \ncline{3c21}{3c22}
  \ncline{3c31}{3c32}
  \ncline{3c32}{3c33}
  
  \ncline{3t}{3l5}
  \ncline{3l5}{3l4}
  \ncline{3t}{3r5}
  \ncline{3r5}{3r4}
  \ncline{3r4}{3r3}
  \ncline{3l4}{3l3}
  \ncline{3c41}{3c42}

  \ncline{2l5}{2c51}
  \ncline{2r5}{2c51}
  \ncline{3l5}{3c51}
  \ncline{3r5}{3c51}

  \ncline{2l4}{2c41}
  \ncline{2c42}{2r4}
  \ncline{3l4}{3c41}
  \ncline{3c42}{3r4}
  
  \ncline{2c33}{3c31}
  
  \ncline{2c22}{1c51}
  \ncline{1c51}{3c21}

  \ncline{2c11}{1c41}
  \ncline{3c11}{1c42}

  \pnode(-9,9){t1l}
  \pnode(-3,9){t1r}
  \rput[bl](-8.5,8.5){\psframebox[framesep=2pt]{$\tau_1(\omega_1)$}}
  \rput[b](-6,8.5){\psframebox[framesep=2pt]{$\tau_1(\omega_2)$}}
  \rput[br](-3.5,8.5){\psframebox[framesep=2pt]{$\tau_1(\omega_3)$}}
  \psline(t1l)(-8.5,9)
  \psline(t1r)(-3.5,9)
  \ncline{t1l}{2l3}

  \pnode(3,9){t2r}
  \rput[bl](-2.5,8.5){\psframebox[framesep=2pt]{$\tau_2(\omega_1)$}}
  \rput[b](0,8.5){\psframebox[framesep=2pt]{$\tau_2(\omega_2)$}}
  \rput[br](2.5,8.5){\psframebox[framesep=2pt]{$\tau_2(\omega_3)$}}
  \psline(t1r)(-2.5,9)
  \psline(t2r)(2.5,9)
  
  \pnode(9,9){t3r}
  \rput[bl](3.5,8.5){\psframebox[framesep=2pt]{$\tau_3(\omega_1)$}}
  \rput[b](6,8.5){\psframebox[framesep=2pt]{$\tau_3(\omega_2)$}}
  \rput[br](8.5,8.5){\psframebox[framesep=2pt]{$\tau_3(\omega_3)$}}
  \psline(t2r)(3.5,9)
  \psline(t3r)(8.5,9)
  \ncline{3r3}{t3r}
  
  \rput(-6,7.5){$c_{123}$}
  \rput(0,7.5){$c_{123}$}
  \rput(6,7.5){$c_{123}$}

\end{pspicture}
  \end{center}
  \caption{Copula vine structure for the uniform parameter distribution.}
  \label{fig:copula_vine}
\end{figure*}
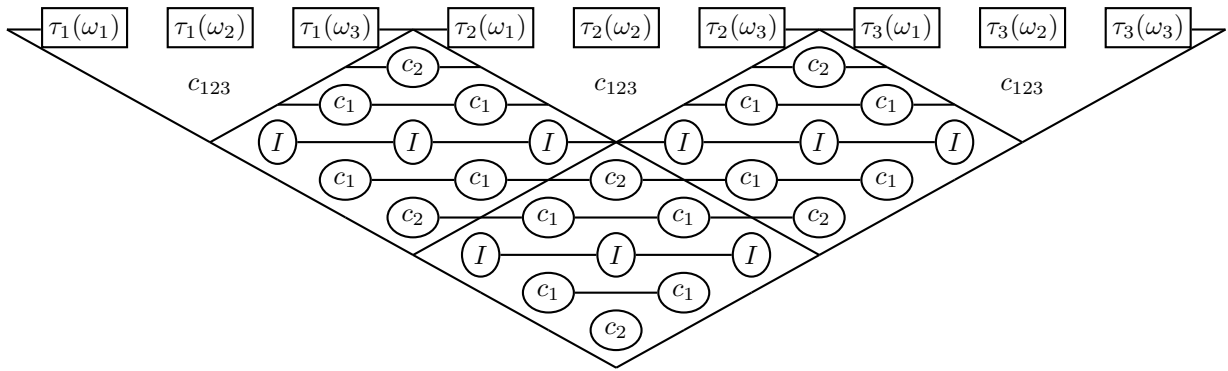

\subsubsection{Target Density for Compensated MCMC}
As the discussion in the previous section suggests, we expect an ensemble of PMD vectors to satisfy the frequency dependent statistical requirements of a true PMD channel when the individual components have a jointly Gaussian distribution.
Since we assume that only the same components at different frequency points are dependent on each other, we can express the joint probability density of a collection of PMD vectors as follows:

\begin{equation}
  f_T(\mathbf{T}, \bar{\tau}) = K(\boldsymbol{\Sigma}) \exp\left(-\frac{1}{2}\mathrm{tr}(\mathbf{T} \boldsymbol{\Sigma}^{-1} \mathbf{T}^T)\right) \quad.
  \label{eqn:mcmc_target}
\end{equation}

\noindent Here, $\mathbf{T}$ is the $3 \times k$ matrix of PMD vectors, $\mathbf{T} = [\boldsymbol{\tau}(\omega_1) \, \boldsymbol{\tau}(\omega_2) \, \dots \, \boldsymbol{\tau}(\omega_k)]$, $\boldsymbol{\Sigma}$ is the $k \times k$ covariance matrix and $K(\boldsymbol{\Sigma})$ is the normalizing constant of the PDF.

The covariance matrix depends on only one parameter which is the desired mean DGD, $\bar{\tau}$, and can be computed with the expression describing the expected value of the inner product of two PMD vectors at different frequency points $\omega$ and $\omega'$ \cite{karlsson1999autocorrelation}:

\begin{equation}
  \left<\boldsymbol{\tau}(\omega) \cdot\boldsymbol{ \tau}(\omega') \right> = \frac{3}{\Delta \omega^2} \left(1 - \exp\left(-\frac{\Delta \omega^2 \left<\bar{\tau}^2\right>}{3}\right) \right) ,
  \label{eqn:pmd_corr}
\end{equation}

\noindent where $\Delta \omega = |\omega-\omega'|$ and $\left<\bar{\tau}^2\right> = \frac{3 \pi}{8}\bar{\tau}^2$.
Using (\ref{eqn:pmd_corr}) we can compute the entries of $\boldsymbol{\Sigma}$ with

\begin{equation*}
  \boldsymbol{\Sigma}_{ij} = \frac{1}{3} \left<\boldsymbol{\tau}(\omega_i) \cdot\boldsymbol{ \tau}(\omega_j) \right> \quad,
\end{equation*} 

\noindent since different components of PMD vectors are independent and since the covariance between two components depends on their distance on the frequency axis, this matrix has a Toeplitz structure.

\subsubsection{Uniform Parameter Distribution of the FIR Filter}

\begin{figure}
  \begin{center}
    \input{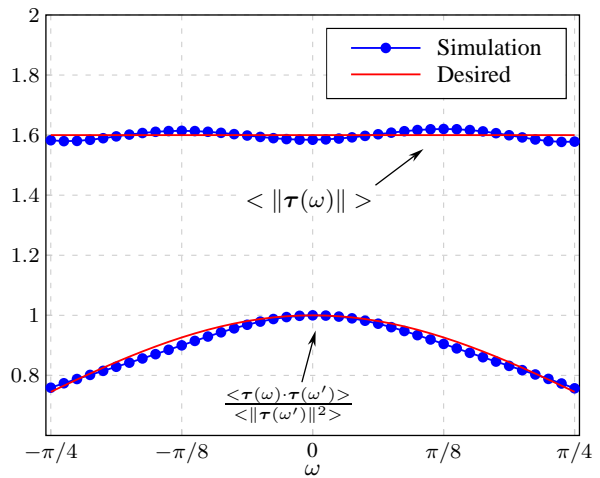}
  \end{center}
  \caption{Mean and normalized autocorrelation curves of the compensated MCMC method compared with the expected values.}
  \label{fig:mcmc_mean_corr}
\end{figure}

\begin{figure*}
  \begin{center}
    \subfigure[PDF at $\omega = 0$]{
%
%
%
\providelength{\AxesLineWidth}       \setlength{\AxesLineWidth}{0.5pt}%
\providelength{\GridLineWidth}       \setlength{\GridLineWidth}{0.4pt}%
\providelength{\GridLineDotSep}      \setlength{\GridLineDotSep}{0.4pt}%
\providelength{\MinorGridLineWidth}  \setlength{\MinorGridLineWidth}{0.4pt}%
\providelength{\MinorGridLineDotSep} \setlength{\MinorGridLineDotSep}{0.8pt}%
\providelength{\plotwidth}           \setlength{\plotwidth}{6cm}
\providelength{\LineWidth}           \setlength{\LineWidth}{0.7pt}%
\providelength{\MarkerSize}          \setlength{\MarkerSize}{2pt}%
\newrgbcolor{GridColor}{0.8 0.8 0.8}%
%
\psset{xunit=0.126273\plotwidth,yunit=0.069715\plotwidth}%
\begin{pspicture}(-0.838710,-11.140351)(7.080645,0.263158)%


\psline[linestyle=dashed,dash=2pt 3pt,dotsep=\GridLineDotSep,linewidth=\GridLineWidth,linecolor=GridColor](0.000000,-10.000000)(0.000000,0.000000)
\psline[linestyle=dashed,dash=2pt 3pt,dotsep=\GridLineDotSep,linewidth=\GridLineWidth,linecolor=GridColor](1.000000,-10.000000)(1.000000,0.000000)
\psline[linestyle=dashed,dash=2pt 3pt,dotsep=\GridLineDotSep,linewidth=\GridLineWidth,linecolor=GridColor](2.000000,-10.000000)(2.000000,0.000000)
\psline[linestyle=dashed,dash=2pt 3pt,dotsep=\GridLineDotSep,linewidth=\GridLineWidth,linecolor=GridColor](3.000000,-10.000000)(3.000000,0.000000)
\psline[linestyle=dashed,dash=2pt 3pt,dotsep=\GridLineDotSep,linewidth=\GridLineWidth,linecolor=GridColor](4.000000,-10.000000)(4.000000,0.000000)
\psline[linestyle=dashed,dash=2pt 3pt,dotsep=\GridLineDotSep,linewidth=\GridLineWidth,linecolor=GridColor](5.000000,-10.000000)(5.000000,0.000000)
\psline[linestyle=dashed,dash=2pt 3pt,dotsep=\GridLineDotSep,linewidth=\GridLineWidth,linecolor=GridColor](6.000000,-10.000000)(6.000000,0.000000)
\psline[linestyle=dashed,dash=2pt 3pt,dotsep=\GridLineDotSep,linewidth=\GridLineWidth,linecolor=GridColor](7.000000,-10.000000)(7.000000,0.000000)
\psline[linestyle=dashed,dash=2pt 3pt,dotsep=\GridLineDotSep,linewidth=\GridLineWidth,linecolor=GridColor](0.000000,-10.000000)(7.000000,-10.000000)
\psline[linestyle=dashed,dash=2pt 3pt,dotsep=\GridLineDotSep,linewidth=\GridLineWidth,linecolor=GridColor](0.000000,-5.000000)(7.000000,-5.000000)
\psline[linestyle=dashed,dash=2pt 3pt,dotsep=\GridLineDotSep,linewidth=\GridLineWidth,linecolor=GridColor](0.000000,0.000000)(7.000000,0.000000)


\psline[linewidth=\AxesLineWidth,linecolor=GridColor](0.000000,-10.000000)(0.000000,-9.847853)
\psline[linewidth=\AxesLineWidth,linecolor=GridColor](1.000000,-10.000000)(1.000000,-9.847853)
\psline[linewidth=\AxesLineWidth,linecolor=GridColor](2.000000,-10.000000)(2.000000,-9.847853)
\psline[linewidth=\AxesLineWidth,linecolor=GridColor](3.000000,-10.000000)(3.000000,-9.847853)
\psline[linewidth=\AxesLineWidth,linecolor=GridColor](4.000000,-10.000000)(4.000000,-9.847853)
\psline[linewidth=\AxesLineWidth,linecolor=GridColor](5.000000,-10.000000)(5.000000,-9.847853)
\psline[linewidth=\AxesLineWidth,linecolor=GridColor](6.000000,-10.000000)(6.000000,-9.847853)
\psline[linewidth=\AxesLineWidth,linecolor=GridColor](7.000000,-10.000000)(7.000000,-9.847853)
\psline[linewidth=\AxesLineWidth,linecolor=GridColor](0.000000,-10.000000)(0.084000,-10.000000)
\psline[linewidth=\AxesLineWidth,linecolor=GridColor](0.000000,-5.000000)(0.084000,-5.000000)
\psline[linewidth=\AxesLineWidth,linecolor=GridColor](0.000000,0.000000)(0.084000,0.000000)


{ \footnotesize 
\rput[t](0.000000,-10.152147){$0$}
\rput[t](1.000000,-10.152147){$1$}
\rput[t](2.000000,-10.152147){$2$}
\rput[t](3.000000,-10.152147){$3$}
\rput[t](4.000000,-10.152147){$4$}
\rput[t](5.000000,-10.152147){$5$}
\rput[t](6.000000,-10.152147){$6$}
\rput[t](7.000000,-10.152147){$7$}
\rput[r](-0.084000,-10.000000){$10^{-10}$}
\rput[r](-0.084000,-5.000000){$10^{-5}$}
\rput[r](-0.084000,0.000000){$10^{0}$}
} 

\psframe[linewidth=\AxesLineWidth,dimen=middle](0.000000,-10.000000)(7.000000,0.000000)

{ \small 
\rput[b](3.500000,-11.5){
\begin{tabular}{c}
$\tau$\\
\end{tabular}
}

\rput[t]{90}(-1.5,-5.000000){
\begin{tabular}{c}
relative frequency\\
\end{tabular}
}
} 

\newrgbcolor{color39.0081}{1  0  0}
\psline[plotstyle=line,linejoin=1,linestyle=solid,linewidth=\LineWidth,linecolor=color39.0081]
(0.050000,-2.704110)(0.150000,-1.754187)(0.250000,-1.319130)(0.350000,-1.039834)(0.450000,-0.838825)
(0.550000,-0.686124)(0.650000,-0.566943)(0.750000,-0.472887)(0.850000,-0.398732)(0.950000,-0.341003)
(1.050000,-0.297272)(1.150000,-0.265775)(1.250000,-0.245190)(1.350000,-0.234503)(1.450000,-0.232914)
(1.550000,-0.239787)(1.650000,-0.254603)(1.750000,-0.276935)(1.850000,-0.306427)(1.950000,-0.342782)
(2.050000,-0.385743)(2.150000,-0.435094)(2.250000,-0.490646)(2.350000,-0.552236)(2.450000,-0.619720)
(2.550000,-0.692972)(2.650000,-0.771880)(2.750000,-0.856347)(2.850000,-0.946283)(2.950000,-1.041609)
(3.050000,-1.142253)(3.150000,-1.248152)(3.250000,-1.359247)(3.350000,-1.475484)(3.450000,-1.596816)
(3.550000,-1.723197)(3.650000,-1.854589)(3.750000,-1.990952)(3.850000,-2.132254)(3.950000,-2.278461)
(4.050000,-2.429545)(4.150000,-2.585480)(4.250000,-2.746238)(4.350000,-2.911798)(4.450000,-3.082137)
(4.550000,-3.257234)(4.650000,-3.437071)(4.750000,-3.621631)(4.850000,-3.810895)(4.950000,-4.004848)
(5.050000,-4.203476)(5.150000,-4.406765)(5.250000,-4.614701)(5.350000,-4.827272)(5.450000,-5.044467)
(5.550000,-5.266275)(5.650000,-5.492684)(5.750000,-5.723686)(5.850000,-5.959270)(5.950000,-6.199428)
(6.050000,-6.444152)(6.150000,-6.693433)(6.250000,-6.947264)(6.350000,-7.205637)(6.450000,-7.468545)
(6.550000,-7.735982)(6.650000,-8.007942)(6.750000,-8.284418)(6.850000,-8.565405)(6.950000,-8.850897)

\newrgbcolor{color40.0071}{0  0  1}
\psline[plotstyle=line,linejoin=1,showpoints=true,dotstyle=*,dotsize=\MarkerSize,linestyle=solid,linewidth=\LineWidth,linecolor=color40.0071]
(0.050000,-2.559422)(0.150000,-1.727182)(0.250000,-1.292968)(0.350000,-1.034650)(0.450000,-0.815742)
(0.550000,-0.672179)(0.650000,-0.551465)(0.750000,-0.464206)(0.850000,-0.392923)(0.950000,-0.334074)
(1.050000,-0.288992)(1.150000,-0.255715)(1.250000,-0.247325)(1.350000,-0.232407)(1.450000,-0.227246)
(1.550000,-0.240949)(1.650000,-0.254038)(1.750000,-0.275035)(1.850000,-0.305797)(1.950000,-0.345074)
(2.050000,-0.389968)(2.150000,-0.440660)(2.250000,-0.497156)(2.350000,-0.562542)(2.450000,-0.634877)
(2.550000,-0.722982)(2.650000,-0.794710)(2.750000,-0.865420)(2.850000,-0.959731)(2.950000,-1.056479)
(3.050000,-1.171917)(3.150000,-1.278363)(3.250000,-1.366478)(3.350000,-1.486082)(3.450000,-1.628446)
(3.550000,-1.766870)(3.650000,-1.884302)(3.750000,-2.022276)(3.850000,-2.225291)(3.950000,-2.367063)
(4.050000,-2.462512)(4.150000,-2.751124)(4.250000,-2.747275)(4.350000,-3.138875)(4.450000,-3.221849)
(4.550000,-3.479413)(4.650000,-3.452679)(4.750000,-3.847390)(4.850000,-3.917026)(4.950000,-4.324511)
\psline[plotstyle=line,linejoin=1,showpoints=true,dotstyle=*,dotsize=\MarkerSize,linestyle=solid,linewidth=\LineWidth,linecolor=color40.0071]
(5.150000,-4.375664)(5.150000,-4.375664)

{ \small 
\rput[bl](0.168000,-9.695706){%
\psframebox[framesep=0pt,linewidth=\AxesLineWidth]{\psframebox*{\begin{tabular}{l}
\Rnode{a1}{\hspace*{0.0ex}} \hspace*{0.7cm} \Rnode{a2}{~~Maxwellian} \\
\Rnode{a3}{\hspace*{0.0ex}} \hspace*{0.7cm} \Rnode{a4}{~~Simulation} \\
\end{tabular}}
\ncline[linestyle=solid,linewidth=\LineWidth,linecolor=color39.0081]{a1}{a2}
\ncline[linestyle=solid,linewidth=\LineWidth,linecolor=color40.0071]{a3}{a4} \ncput{\psdot[dotstyle=*,dotsize=\MarkerSize,linecolor=color40.0071]}
}%
}%
} 

\end{pspicture}%
       \label{fig:mcmc_pdf_center}
     }
     \subfigure[PDF at $\omega = \frac{\pi}{4}$]{
%
%
%
\providelength{\AxesLineWidth}       \setlength{\AxesLineWidth}{0.5pt}%
\providelength{\GridLineWidth}       \setlength{\GridLineWidth}{0.4pt}%
\providelength{\GridLineDotSep}      \setlength{\GridLineDotSep}{0.4pt}%
\providelength{\MinorGridLineWidth}  \setlength{\MinorGridLineWidth}{0.4pt}%
\providelength{\MinorGridLineDotSep} \setlength{\MinorGridLineDotSep}{0.8pt}%
\providelength{\plotwidth}           \setlength{\plotwidth}{6cm}
\providelength{\LineWidth}           \setlength{\LineWidth}{0.7pt}%
\providelength{\MarkerSize}          \setlength{\MarkerSize}{2pt}%
\newrgbcolor{GridColor}{0.8 0.8 0.8}%
%
\psset{xunit=0.126273\plotwidth,yunit=0.069715\plotwidth}%
\begin{pspicture}(-0.838710,-11.140351)(7.080645,0.263158)%


\psline[linestyle=dashed,dash=2pt 3pt,dotsep=\GridLineDotSep,linewidth=\GridLineWidth,linecolor=GridColor](0.000000,-10.000000)(0.000000,0.000000)
\psline[linestyle=dashed,dash=2pt 3pt,dotsep=\GridLineDotSep,linewidth=\GridLineWidth,linecolor=GridColor](1.000000,-10.000000)(1.000000,0.000000)
\psline[linestyle=dashed,dash=2pt 3pt,dotsep=\GridLineDotSep,linewidth=\GridLineWidth,linecolor=GridColor](2.000000,-10.000000)(2.000000,0.000000)
\psline[linestyle=dashed,dash=2pt 3pt,dotsep=\GridLineDotSep,linewidth=\GridLineWidth,linecolor=GridColor](3.000000,-10.000000)(3.000000,0.000000)
\psline[linestyle=dashed,dash=2pt 3pt,dotsep=\GridLineDotSep,linewidth=\GridLineWidth,linecolor=GridColor](4.000000,-10.000000)(4.000000,0.000000)
\psline[linestyle=dashed,dash=2pt 3pt,dotsep=\GridLineDotSep,linewidth=\GridLineWidth,linecolor=GridColor](5.000000,-10.000000)(5.000000,0.000000)
\psline[linestyle=dashed,dash=2pt 3pt,dotsep=\GridLineDotSep,linewidth=\GridLineWidth,linecolor=GridColor](6.000000,-10.000000)(6.000000,0.000000)
\psline[linestyle=dashed,dash=2pt 3pt,dotsep=\GridLineDotSep,linewidth=\GridLineWidth,linecolor=GridColor](7.000000,-10.000000)(7.000000,0.000000)
\psline[linestyle=dashed,dash=2pt 3pt,dotsep=\GridLineDotSep,linewidth=\GridLineWidth,linecolor=GridColor](0.000000,-10.000000)(7.000000,-10.000000)
\psline[linestyle=dashed,dash=2pt 3pt,dotsep=\GridLineDotSep,linewidth=\GridLineWidth,linecolor=GridColor](0.000000,-5.000000)(7.000000,-5.000000)
\psline[linestyle=dashed,dash=2pt 3pt,dotsep=\GridLineDotSep,linewidth=\GridLineWidth,linecolor=GridColor](0.000000,0.000000)(7.000000,0.000000)


\psline[linewidth=\AxesLineWidth,linecolor=GridColor](0.000000,-10.000000)(0.000000,-9.847853)
\psline[linewidth=\AxesLineWidth,linecolor=GridColor](1.000000,-10.000000)(1.000000,-9.847853)
\psline[linewidth=\AxesLineWidth,linecolor=GridColor](2.000000,-10.000000)(2.000000,-9.847853)
\psline[linewidth=\AxesLineWidth,linecolor=GridColor](3.000000,-10.000000)(3.000000,-9.847853)
\psline[linewidth=\AxesLineWidth,linecolor=GridColor](4.000000,-10.000000)(4.000000,-9.847853)
\psline[linewidth=\AxesLineWidth,linecolor=GridColor](5.000000,-10.000000)(5.000000,-9.847853)
\psline[linewidth=\AxesLineWidth,linecolor=GridColor](6.000000,-10.000000)(6.000000,-9.847853)
\psline[linewidth=\AxesLineWidth,linecolor=GridColor](7.000000,-10.000000)(7.000000,-9.847853)
\psline[linewidth=\AxesLineWidth,linecolor=GridColor](0.000000,-10.000000)(0.084000,-10.000000)
\psline[linewidth=\AxesLineWidth,linecolor=GridColor](0.000000,-5.000000)(0.084000,-5.000000)
\psline[linewidth=\AxesLineWidth,linecolor=GridColor](0.000000,0.000000)(0.084000,0.000000)


{ \footnotesize 
\rput[t](0.000000,-10.152147){$0$}
\rput[t](1.000000,-10.152147){$1$}
\rput[t](2.000000,-10.152147){$2$}
\rput[t](3.000000,-10.152147){$3$}
\rput[t](4.000000,-10.152147){$4$}
\rput[t](5.000000,-10.152147){$5$}
\rput[t](6.000000,-10.152147){$6$}
\rput[t](7.000000,-10.152147){$7$}
\rput[r](-0.084000,-10.000000){$10^{-10}$}
\rput[r](-0.084000,-5.000000){$10^{-5}$}
\rput[r](-0.084000,0.000000){$10^{0}$}
} 

\psframe[linewidth=\AxesLineWidth,dimen=middle](0.000000,-10.000000)(7.000000,0.000000)

{ \small 
\rput[b](3.500000,-11.5){
\begin{tabular}{c}
$\tau$\\
\end{tabular}
}

\rput[t]{90}(-1.5,-5.000000){
\begin{tabular}{c}
relative frequency\\
\end{tabular}
}
} 

\newrgbcolor{color39.0089}{1  0  0}
\psline[plotstyle=line,linejoin=1,linestyle=solid,linewidth=\LineWidth,linecolor=color39.0089]
(0.050000,-2.704110)(0.150000,-1.754187)(0.250000,-1.319130)(0.350000,-1.039834)(0.450000,-0.838825)
(0.550000,-0.686124)(0.650000,-0.566943)(0.750000,-0.472887)(0.850000,-0.398732)(0.950000,-0.341003)
(1.050000,-0.297272)(1.150000,-0.265775)(1.250000,-0.245190)(1.350000,-0.234503)(1.450000,-0.232914)
(1.550000,-0.239787)(1.650000,-0.254603)(1.750000,-0.276935)(1.850000,-0.306427)(1.950000,-0.342782)
(2.050000,-0.385743)(2.150000,-0.435094)(2.250000,-0.490646)(2.350000,-0.552236)(2.450000,-0.619720)
(2.550000,-0.692972)(2.650000,-0.771880)(2.750000,-0.856347)(2.850000,-0.946283)(2.950000,-1.041609)
(3.050000,-1.142253)(3.150000,-1.248152)(3.250000,-1.359247)(3.350000,-1.475484)(3.450000,-1.596816)
(3.550000,-1.723197)(3.650000,-1.854589)(3.750000,-1.990952)(3.850000,-2.132254)(3.950000,-2.278461)
(4.050000,-2.429545)(4.150000,-2.585480)(4.250000,-2.746238)(4.350000,-2.911798)(4.450000,-3.082137)
(4.550000,-3.257234)(4.650000,-3.437071)(4.750000,-3.621631)(4.850000,-3.810895)(4.950000,-4.004848)
(5.050000,-4.203476)(5.150000,-4.406765)(5.250000,-4.614701)(5.350000,-4.827272)(5.450000,-5.044467)
(5.550000,-5.266275)(5.650000,-5.492684)(5.750000,-5.723686)(5.850000,-5.959270)(5.950000,-6.199428)
(6.050000,-6.444152)(6.150000,-6.693433)(6.250000,-6.947264)(6.350000,-7.205637)(6.450000,-7.468545)
(6.550000,-7.735982)(6.650000,-8.007942)(6.750000,-8.284418)(6.850000,-8.565405)(6.950000,-8.850897)

\newrgbcolor{color40.0084}{0  0  1}
\psline[plotstyle=line,linejoin=1,showpoints=true,dotstyle=*,dotsize=\MarkerSize,linestyle=solid,linewidth=\LineWidth,linecolor=color40.0084]
(0.050000,-2.529017)(0.150000,-1.697715)(0.250000,-1.296482)(0.350000,-1.027383)(0.450000,-0.815413)
(0.550000,-0.664690)(0.650000,-0.547858)(0.750000,-0.452802)(0.850000,-0.384415)(0.950000,-0.325284)
(1.050000,-0.284498)(1.150000,-0.258169)(1.250000,-0.235218)(1.350000,-0.232563)(1.450000,-0.233290)
(1.550000,-0.239645)(1.650000,-0.258160)(1.750000,-0.281817)(1.850000,-0.320347)(1.950000,-0.355956)
(2.050000,-0.389665)(2.150000,-0.450474)(2.250000,-0.502648)(2.350000,-0.564600)(2.450000,-0.637795)
(2.550000,-0.708655)(2.650000,-0.787812)(2.750000,-0.871142)(2.850000,-0.959523)(2.950000,-1.067019)
(3.050000,-1.159894)(3.150000,-1.280189)(3.250000,-1.396229)(3.350000,-1.486152)(3.450000,-1.636487)
(3.550000,-1.773604)(3.650000,-1.866629)(3.750000,-1.987607)(3.850000,-2.152297)(3.950000,-2.303322)
(4.050000,-2.395092)(4.150000,-2.601147)(4.250000,-2.743459)(4.350000,-2.920819)(4.450000,-3.091233)
(4.550000,-3.354474)(4.650000,-3.472574)(4.750000,-3.710552)(4.850000,-3.645285)(4.950000,-4.237361)
(5.050000,-4.164810)(5.150000,-4.579784)(5.250000,-4.977724)
\psline[plotstyle=line,linejoin=1,showpoints=true,dotstyle=*,dotsize=\MarkerSize,linestyle=solid,linewidth=\LineWidth,linecolor=color40.0084]
(5.550000,-5.278754)(5.550000,-5.278754)
\psline[plotstyle=line,linejoin=1,showpoints=true,dotstyle=*,dotsize=\MarkerSize,linestyle=solid,linewidth=\LineWidth,linecolor=color40.0084]
(5.750000,-5.278754)(5.750000,-5.278754)

{ \small 
\rput[bl](0.168000,-9.695706){%
\psframebox[framesep=0pt,linewidth=\AxesLineWidth]{\psframebox*{\begin{tabular}{l}
\Rnode{a1}{\hspace*{0.0ex}} \hspace*{0.7cm} \Rnode{a2}{~~Maxwellian} \\
\Rnode{a3}{\hspace*{0.0ex}} \hspace*{0.7cm} \Rnode{a4}{~~Simulation} \\
\end{tabular}}
\ncline[linestyle=solid,linewidth=\LineWidth,linecolor=color39.0089]{a1}{a2}
\ncline[linestyle=solid,linewidth=\LineWidth,linecolor=color40.0084]{a3}{a4} \ncput{\psdot[dotstyle=*,dotsize=\MarkerSize,linecolor=color40.0084]}
}%
}%
} 

\end{pspicture}%
       \label{fig:mcmc_pdf_corner}
     }
  \end{center}
  \caption{PDF of the DGD at the center frequency ($\omega = 0$) and the corner frequency ($\omega = \frac{\pi}{4}$) for compensated MCMC method. Solid curves represent the Maxwellian with mean 1.6.}
  \label{fig:mcmc_dgd_pdfs}
\end{figure*}
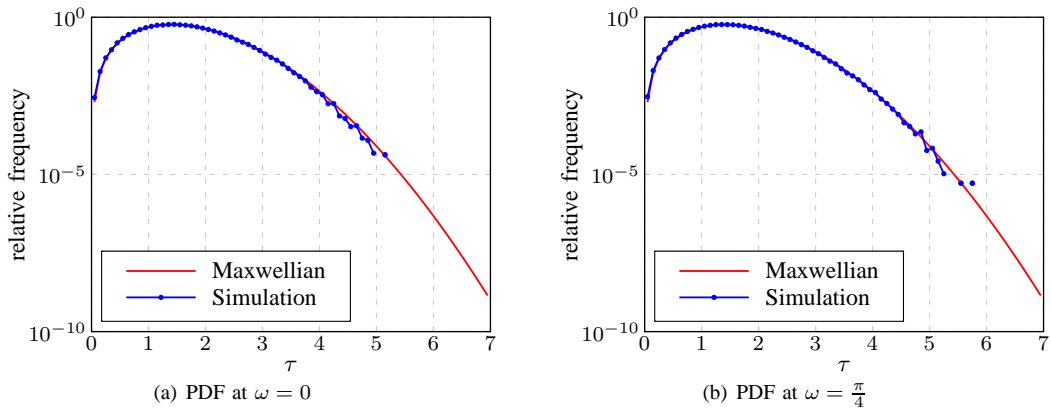

The evaluation of the accept-reject rule in the compensated MCMC algorithm requires the knowledge of the distribution of the PMD vectors when the model parameters are sampled uniformly and independently. 
Although this distribution resembles the PMD vector distribution of the full model, its exact form deviates from a jointly Gaussian distribution much more than the PDF of the full model does because it has significantly fewer birefringent sections.

Indeed, the effort to approximate this PDF with a joint Gaussian or even a Gaussian mixture model results in inaccurate PMD vector statistics in terms of mean DGD and covariance structure. 
Moreover, upon close inspection, one can observe that despite being uncorrelated, different PMD vector components at different frequency points exhibit a tail dependency which cannot be captured with a jointly Gaussian distribution. In order to overcome these difficulties, we model the uniform parameter distribution as a copula vine \cite{bedford2002vines}. 

Copulas are multivariate functions that are employed to describe dependency structures of random variables \cite{nelsen2006introduction, frees1998understanding}. 
For our purposes, without going into details, copulas can be viewed as linkage functions that express the joint PDFs of random variables as the product of their marginals and their dependency structure. 
For the case of two dependent random variables, $X$ and $Y$ with marginal CDFs $F_X$ and $F_Y$, one can write

\begin{equation}
  f_{XY}(x,y) = f_X(x) f_Y(y) c_{XY}(F_X(x), F_Y(y)) \quad,
  \label{eqn:copula_decomp}
\end{equation}

\noindent where $f_X$ and $f_Y$ are marginal PDFs of $X$ and $Y$ respectively and $f_{XY}$ is their joint PDF. 
The non-negative bivariate function $c_{XY}: [0,1] \times [0,1] \to \mathbb{R}_+$ is the copula density of the two random variables. According to Sklar's theorem \cite{sklar1959fonctions}, under some regularity conditions, such a function always exists.
Equation (\ref{eqn:copula_decomp}) becomes especially useful if the marginals are known. This is indeed the case in the UPD. The PMD vector at the i$^\text{th}$ frequency point, $\boldsymbol{\tau}(\omega_i) = [\tau_1(\omega_i), \tau_2(\omega_i), \tau_3(\omega_i)]^T$, has independent components that are distributed according to the uniform sum distribution. A random variable $X$ that is the sum of $N$ independent uniform random variables has the PDF

\begin{equation}
  f_U(x,N)=\frac{1}{2\left(N-1\right)!}\sum_{k=0}^{N}\left(-1\right)^k{N \choose k}\left(x-k\right)^{N-1}\text{sgn}(x-k) .
  \label{eqn:uni_sum_dist_pdf}
\end{equation}

\noindent Its CDF is given by

\begin{equation}
  F_{U}(x,N) = \frac{1}{N!}\sum_{k=0}^{\lfloor x\rfloor}(-1)^k\binom{N}{k}(x-k)^N \quad.
  \label{eqn:uni_sum_dist_cdf}
\end{equation}

\noindent The subscript $U$ denotes that this distribution is the univariate marginal of the \underline{U}PD and N is the number of sections of the paraunitary FIR filter. 

For the construction of the copula density, the multivariate t-copula has proven to be useful \cite{demarta2005t}. 
In fact, the set of same PMD vector components (e.g. the first component at all the frequency points) follows a multivariate t-copula distribution with uniform sum marginals almost exactly. 
The remaining dependency among cross-components is modeled using bivariate t-copulas and arranging them into a D-vine \cite{aas2009pair} in order to exploit the stationarity property of the joint PMD vector distribution. 
The parameters of the multivariate as well as the bivariate copulas can be estimated using standard maximum likelihood algorithms.

Figure \ref{fig:copula_vine} illustrates the strategy for building trees of pair copulas for three frequency points. 
The three-variate t-copula, $c_{123}(F_{U}(\tau_m(\omega_1)), F_{U}(\tau_m(\omega_2)), F_{U}(\tau_m(\omega_3)))$, is the joint density of the $m^\text{th}$ PMD vector components. 
The dependency among cross-components, $(\tau_m(\omega_i), \tau_n(\omega_j)), \; m \neq n,\; i \neq j$, is accounted for with the pair copulas $c_1$ and $c_2$, and $I$ is the independence copula that connects the components of a single PMD vector. 
The symmetry in Figure \ref{fig:copula_vine} is due to a simplifying assumption we make in order to obtain a more tractable UPD. 
Note that the pair copulas connecting the cross-components are in fact conditional PDFs that not only depend on the conditioning variables but also operate on the transformed forms of their arguments. 
Here we make the assumption that these copulas are sufficiently flat so that we can describe the relationship among cross-components solely based on their frequency separation. 
In the end, we obtain the whole joint copula of all nine variables by multiplying all the components in Figure \ref{fig:copula_vine}.

\subsubsection{Performance of Compensated MCMC}

Based on the above, we can construct an accept-reject rule in the compensated MCMC algorithm such that the statistics of the ensemble of output PMD vectors will approximate the desired values. To this end, based on (\ref{eq:mod-accpt-rejct}),  we can write

\begin{equation}
  \alpha_c = \min \left(1, \frac{f_{T}(\tilde{\mathbf{T}}, \bar{\tau})}{f_{T}(\mathbf{T}, \bar{\tau})} \frac{f_{U}(\mathbf{T},N)}{f_{U}(\tilde{\mathbf{T}}, N)}\right) \quad,
  \label{eqn:comp-accpt-rej}
\end{equation}

The statistics of the C-MCMC algorithm output is illustrated in Figures \ref{fig:mcmc_mean_corr} and \ref{fig:mcmc_dgd_pdfs}. The simulation was ran with $2 \times 10^6$ samples and the first $10^4$ samples were discarded as the burn-in phase. The number of frequencies in the accept-reject rule, $k$, was chosen to be 3 ($-\pi/4$, $0$ and $\pi/4$). It can be observed that this algorithm performs much better in terms of the autocorrelation function than the cascade algorithm at the cost of a small deviation in the mean value.

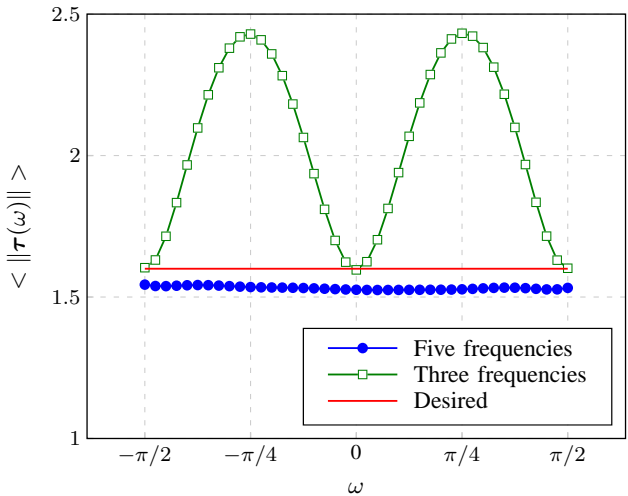
\begin{figure}
  \begin{center}
%
%
%
\providelength{\AxesLineWidth}       \setlength{\AxesLineWidth}{0.5pt}%
\providelength{\GridLineWidth}       \setlength{\GridLineWidth}{0.4pt}%
\providelength{\GridLineDotSep}      \setlength{\GridLineDotSep}{0.4pt}%
\providelength{\MinorGridLineWidth}  \setlength{\MinorGridLineWidth}{0.4pt}%
\providelength{\MinorGridLineDotSep} \setlength{\MinorGridLineDotSep}{0.8pt}%
\providelength{\plotwidth}           \setlength{\plotwidth}{8cm}
\providelength{\LineWidth}           \setlength{\LineWidth}{0.7pt}%
\providelength{\MarkerSize}          \setlength{\MarkerSize}{4pt}%
\newrgbcolor{GridColor}{0.8 0.8 0.8}%
%
\psset{xunit=0.223711\plotwidth,yunit=0.470515\plotwidth}%
\begin{pspicture}(-2.451613,0.828947)(2.018433,2.539474)%


\psline[linestyle=dashed,dash=2pt 3pt,dotsep=\GridLineDotSep,linewidth=\GridLineWidth,linecolor=GridColor](-1.570796,1.000000)(-1.570796,2.500000)
\psline[linestyle=dashed,dash=2pt 3pt,dotsep=\GridLineDotSep,linewidth=\GridLineWidth,linecolor=GridColor](-0.785398,1.000000)(-0.785398,2.500000)
\psline[linestyle=dashed,dash=2pt 3pt,dotsep=\GridLineDotSep,linewidth=\GridLineWidth,linecolor=GridColor](0.000000,1.000000)(0.000000,2.500000)
\psline[linestyle=dashed,dash=2pt 3pt,dotsep=\GridLineDotSep,linewidth=\GridLineWidth,linecolor=GridColor](0.785398,1.000000)(0.785398,2.500000)
\psline[linestyle=dashed,dash=2pt 3pt,dotsep=\GridLineDotSep,linewidth=\GridLineWidth,linecolor=GridColor](1.570796,1.000000)(1.570796,2.500000)
\psline[linestyle=dashed,dash=2pt 3pt,dotsep=\GridLineDotSep,linewidth=\GridLineWidth,linecolor=GridColor](-2.000000,1.000000)(2.000000,1.000000)
\psline[linestyle=dashed,dash=2pt 3pt,dotsep=\GridLineDotSep,linewidth=\GridLineWidth,linecolor=GridColor](-2.000000,1.500000)(2.000000,1.500000)
\psline[linestyle=dashed,dash=2pt 3pt,dotsep=\GridLineDotSep,linewidth=\GridLineWidth,linecolor=GridColor](-2.000000,2.000000)(2.000000,2.000000)
\psline[linestyle=dashed,dash=2pt 3pt,dotsep=\GridLineDotSep,linewidth=\GridLineWidth,linecolor=GridColor](-2.000000,2.500000)(2.000000,2.500000)

\psline[linewidth=\AxesLineWidth,linecolor=GridColor](-1.570796,1.000000)(-1.570796,1.022822)
\psline[linewidth=\AxesLineWidth,linecolor=GridColor](-0.785398,1.000000)(-0.785398,1.022822)
\psline[linewidth=\AxesLineWidth,linecolor=GridColor](0.000000,1.000000)(0.000000,1.022822)
\psline[linewidth=\AxesLineWidth,linecolor=GridColor](0.785398,1.000000)(0.785398,1.022822)
\psline[linewidth=\AxesLineWidth,linecolor=GridColor](1.570796,1.000000)(1.570796,1.022822)
\psline[linewidth=\AxesLineWidth,linecolor=GridColor](-2.000000,1.000000)(-1.952000,1.000000)
\psline[linewidth=\AxesLineWidth,linecolor=GridColor](-2.000000,1.500000)(-1.952000,1.500000)
\psline[linewidth=\AxesLineWidth,linecolor=GridColor](-2.000000,2.000000)(-1.952000,2.000000)
\psline[linewidth=\AxesLineWidth,linecolor=GridColor](-2.000000,2.500000)(-1.952000,2.500000)

{ \footnotesize 
\rput[t](-1.570796,0.977178){$-\pi/2$}
\rput[t](-0.785398,0.977178){$-\pi/4$}
\rput[t](0.000000,0.977178){0}
\rput[t](0.785398,0.977178){$\pi/4$}
\rput[t](1.570796,0.977178){$\pi/2$}
\rput[r](-2.048000,1.000000){$1$}
\rput[r](-2.048000,1.500000){$1.5$}
\rput[r](-2.048000,2.000000){$2$}
\rput[r](-2.048000,2.500000){$2.5$}
} 

\psframe[linewidth=\AxesLineWidth,dimen=middle](-2.000000,1.000000)(2.000000,2.500000)

{ \small 
\rput[b](0.000000,0.78){
\begin{tabular}{c}
$\omega$\\
\end{tabular}
}

\rput[t]{90}(-2.6,1.750000){
\begin{tabular}{c}
$<\|\boldsymbol{\tau}(\omega)\|>$\\
\end{tabular}
}
} 

\newrgbcolor{color344.0219}{0  0  1}
\psline[plotstyle=line,linejoin=1,showpoints=true,dotstyle=*,dotsize=\MarkerSize,linestyle=solid,linewidth=\LineWidth,linecolor=color344.0219]
(-1.570796,1.543819)(-1.492257,1.538910)(-1.413717,1.538455)(-1.335177,1.539936)(-1.256637,1.541582)
(-1.178097,1.542328)(-1.099557,1.541861)(-1.021018,1.540427)(-0.942478,1.538512)(-0.863938,1.536666)
(-0.785398,1.535238)(-0.706858,1.534301)(-0.628319,1.533733)(-0.549779,1.533299)(-0.471239,1.532721)
(-0.392699,1.531827)(-0.314159,1.530653)(-0.235619,1.529312)(-0.157080,1.527939)(-0.078540,1.526677)
(0.000000,1.525677)(0.078540,1.525054)(0.157080,1.524819)(0.235619,1.524888)(0.314159,1.525114)
(0.392699,1.525356)(0.471239,1.525526)(0.549779,1.525645)(0.628319,1.525838)(0.706858,1.526315)
(0.785398,1.527290)(0.863938,1.528810)(0.942478,1.530626)(1.021018,1.532265)(1.099557,1.533180)
(1.178097,1.532914)(1.256637,1.531361)(1.335177,1.528971)(1.413717,1.526917)(1.492257,1.527125)
(1.570796,1.532308)

\newrgbcolor{color345.0209}{0     0.49804           0}
\psline[plotstyle=line,linejoin=1,showpoints=true,dotstyle=Bsquare,dotsize=\MarkerSize,linestyle=solid,linewidth=\LineWidth,linecolor=color345.0209]
(-1.570796,1.603662)(-1.492257,1.630930)(-1.413717,1.714733)(-1.335177,1.833679)(-1.256637,1.966793)
(-1.178097,2.097681)(-1.099557,2.214789)(-1.021018,2.310397)(-0.942478,2.379613)(-0.863938,2.419707)
(-0.785398,2.429505)(-0.706858,2.409040)(-0.628319,2.359273)(-0.549779,2.282380)(-0.471239,2.182071)
(-0.392699,2.063950)(-0.314159,1.936146)(-0.235619,1.809943)(-0.157080,1.700004)(-0.078540,1.623308)
(0.000000,1.595948)(0.078540,1.624837)(0.157080,1.702529)(0.235619,1.813090)(0.314159,1.939824)
(0.392699,2.068001)(0.471239,2.186230)(0.549779,2.286395)(0.628319,2.362911)(0.706858,2.412235)
(0.785398,2.432299)(0.863938,2.422123)(0.942478,2.381779)(1.021018,2.312454)(1.099557,2.216776)
(1.178097,2.099500)(1.256637,1.968341)(1.335177,1.834810)(1.413717,1.715175)(1.492257,1.630364)
(1.570796,1.601736)

\newrgbcolor{color346.0209}{1  0  0}
\psline[plotstyle=line,linejoin=1,linestyle=solid,linewidth=\LineWidth,linecolor=color346.0209]
(-1.570796,1.600000)(-1.492257,1.600000)(-1.413717,1.600000)(-1.335177,1.600000)(-1.256637,1.600000)
(-1.178097,1.600000)(-1.099557,1.600000)(-1.021018,1.600000)(-0.942478,1.600000)(-0.863938,1.600000)
(-0.785398,1.600000)(-0.706858,1.600000)(-0.628319,1.600000)(-0.549779,1.600000)(-0.471239,1.600000)
(-0.392699,1.600000)(-0.314159,1.600000)(-0.235619,1.600000)(-0.157080,1.600000)(-0.078540,1.600000)
(0.000000,1.600000)(0.078540,1.600000)(0.157080,1.600000)(0.235619,1.600000)(0.314159,1.600000)
(0.392699,1.600000)(0.471239,1.600000)(0.549779,1.600000)(0.628319,1.600000)(0.706858,1.600000)
(0.785398,1.600000)(0.863938,1.600000)(0.942478,1.600000)(1.021018,1.600000)(1.099557,1.600000)
(1.178097,1.600000)(1.256637,1.600000)(1.335177,1.600000)(1.413717,1.600000)(1.492257,1.600000)
(1.570796,1.600000)

{ \small 
\rput[br](1.904000,1.045644){%
\psframebox[framesep=0pt,linewidth=\AxesLineWidth]{\psframebox*{\begin{tabular}{l}
\Rnode{a1}{\hspace*{0.0ex}} \hspace*{0.7cm} \Rnode{a2}{~~Five frequencies} \\
\Rnode{a3}{\hspace*{0.0ex}} \hspace*{0.7cm} \Rnode{a4}{~~Three frequencies} \\
\Rnode{a5}{\hspace*{0.0ex}} \hspace*{0.7cm} \Rnode{a6}{~~Desired} \\
\end{tabular}}
\ncline[linestyle=solid,linewidth=\LineWidth,linecolor=color344.0219]{a1}{a2} \ncput{\psdot[dotstyle=*,dotsize=\MarkerSize,linecolor=color344.0219]}
\ncline[linestyle=solid,linewidth=\LineWidth,linecolor=color345.0209]{a3}{a4} \ncput{\psdot[dotstyle=Bsquare,dotsize=\MarkerSize,linecolor=color345.0209]}
\ncline[linestyle=solid,linewidth=\LineWidth,linecolor=color346.0209]{a5}{a6}
}%
}%
} 

\end{pspicture}%
  \end{center}
  \caption{The mean DGD values for a twice oversampled system with three and five frequency points in the accept-reject rule of the compensated MCMC algorithm.}
  \label{fig:mcmc_three_five_freq_mean}
\end{figure}

The limiting factor on the accuracy of the C-MCMC algorithm is how well the UDP can be modeled and approximated.
The simplifying assumptions about the UDP and the pair copulas describing it are sources of  inaccuracy. As the number of frequencies in the accept-reject rule, and consequently the number of pair copulas used to approximate the UDP grows, the approximation becomes increasingly inaccurate and the PDFs do not converge to the desired curves.
On the other hand, holding the number of frequencies fixed while increasing their distance results in non-uniform frequency behavior of the output. 
Figure \ref{fig:mcmc_three_five_freq_mean} illustrates these two properties together by displaying the mean DGD values for simulations using three and five frequencies over the frequency range $[-\pi/2,\pi/2]$, i.e.\ for a system that is oversampled at twice its minimum sampling rate. The simulation with three frequency points in the accept-reject rule matches the desired mean DGD value at these frequencies but exhibits large deviations at intermediate points while the simulation with five frequencies results in a relatively constant mean DGD value smaller than the desired one.

\subsection{Greedy Transfer Function Approximation Method }

\begin{figure}
  \begin{center}
    \input{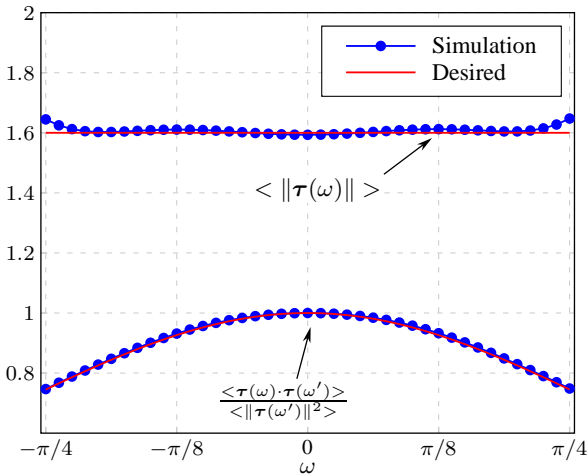}
  \end{center}
  \caption{Mean and normalized autocorrelation curves of the greedy approximation method compared with the expected values.}
  \label{fig:greedy_mean_corr}
\end{figure}

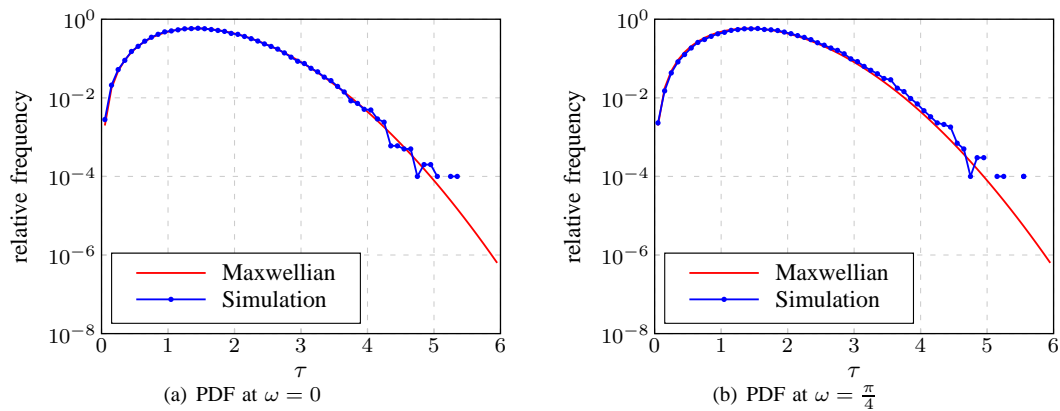
\begin{figure*}
  \begin{center}
    \subfigure[PDF at $\omega = 0$]{
%
%
%
\providelength{\AxesLineWidth}       \setlength{\AxesLineWidth}{0.5pt}%
\providelength{\GridLineWidth}       \setlength{\GridLineWidth}{0.4pt}%
\providelength{\GridLineDotSep}      \setlength{\GridLineDotSep}{0.4pt}%
\providelength{\MinorGridLineWidth}  \setlength{\MinorGridLineWidth}{0.4pt}%
\providelength{\MinorGridLineDotSep} \setlength{\MinorGridLineDotSep}{0.8pt}%
\providelength{\plotwidth}           \setlength{\plotwidth}{6cm}
\providelength{\LineWidth}           \setlength{\LineWidth}{0.7pt}%
\providelength{\MarkerSize}          \setlength{\MarkerSize}{2pt}%
\newrgbcolor{GridColor}{0.8 0.8 0.8}%
%
\psset{xunit=0.147318\plotwidth,yunit=0.087144\plotwidth}%
\begin{pspicture}(-0.718894,-8.912281)(6.069124,0.210526)%


\psline[linestyle=dashed,dash=2pt 3pt,dotsep=\GridLineDotSep,linewidth=\GridLineWidth,linecolor=GridColor](0.000000,-8.000000)(0.000000,0.000000)
\psline[linestyle=dashed,dash=2pt 3pt,dotsep=\GridLineDotSep,linewidth=\GridLineWidth,linecolor=GridColor](1.000000,-8.000000)(1.000000,0.000000)
\psline[linestyle=dashed,dash=2pt 3pt,dotsep=\GridLineDotSep,linewidth=\GridLineWidth,linecolor=GridColor](2.000000,-8.000000)(2.000000,0.000000)
\psline[linestyle=dashed,dash=2pt 3pt,dotsep=\GridLineDotSep,linewidth=\GridLineWidth,linecolor=GridColor](3.000000,-8.000000)(3.000000,0.000000)
\psline[linestyle=dashed,dash=2pt 3pt,dotsep=\GridLineDotSep,linewidth=\GridLineWidth,linecolor=GridColor](4.000000,-8.000000)(4.000000,0.000000)
\psline[linestyle=dashed,dash=2pt 3pt,dotsep=\GridLineDotSep,linewidth=\GridLineWidth,linecolor=GridColor](5.000000,-8.000000)(5.000000,0.000000)
\psline[linestyle=dashed,dash=2pt 3pt,dotsep=\GridLineDotSep,linewidth=\GridLineWidth,linecolor=GridColor](6.000000,-8.000000)(6.000000,0.000000)
\psline[linestyle=dashed,dash=2pt 3pt,dotsep=\GridLineDotSep,linewidth=\GridLineWidth,linecolor=GridColor](0.000000,-8.000000)(6.000000,-8.000000)
\psline[linestyle=dashed,dash=2pt 3pt,dotsep=\GridLineDotSep,linewidth=\GridLineWidth,linecolor=GridColor](0.000000,-6.000000)(6.000000,-6.000000)
\psline[linestyle=dashed,dash=2pt 3pt,dotsep=\GridLineDotSep,linewidth=\GridLineWidth,linecolor=GridColor](0.000000,-4.000000)(6.000000,-4.000000)
\psline[linestyle=dashed,dash=2pt 3pt,dotsep=\GridLineDotSep,linewidth=\GridLineWidth,linecolor=GridColor](0.000000,-2.000000)(6.000000,-2.000000)
\psline[linestyle=dashed,dash=2pt 3pt,dotsep=\GridLineDotSep,linewidth=\GridLineWidth,linecolor=GridColor](0.000000,0.000000)(6.000000,0.000000)


\psline[linewidth=\AxesLineWidth,linecolor=GridColor](0.000000,-8.000000)(0.000000,-7.878282)
\psline[linewidth=\AxesLineWidth,linecolor=GridColor](1.000000,-8.000000)(1.000000,-7.878282)
\psline[linewidth=\AxesLineWidth,linecolor=GridColor](2.000000,-8.000000)(2.000000,-7.878282)
\psline[linewidth=\AxesLineWidth,linecolor=GridColor](3.000000,-8.000000)(3.000000,-7.878282)
\psline[linewidth=\AxesLineWidth,linecolor=GridColor](4.000000,-8.000000)(4.000000,-7.878282)
\psline[linewidth=\AxesLineWidth,linecolor=GridColor](5.000000,-8.000000)(5.000000,-7.878282)
\psline[linewidth=\AxesLineWidth,linecolor=GridColor](6.000000,-8.000000)(6.000000,-7.878282)
\psline[linewidth=\AxesLineWidth,linecolor=GridColor](0.000000,-8.000000)(0.072000,-8.000000)
\psline[linewidth=\AxesLineWidth,linecolor=GridColor](0.000000,-6.000000)(0.072000,-6.000000)
\psline[linewidth=\AxesLineWidth,linecolor=GridColor](0.000000,-4.000000)(0.072000,-4.000000)
\psline[linewidth=\AxesLineWidth,linecolor=GridColor](0.000000,-2.000000)(0.072000,-2.000000)
\psline[linewidth=\AxesLineWidth,linecolor=GridColor](0.000000,0.000000)(0.072000,0.000000)


{ \footnotesize 
\rput[t](0.000000,-8.121718){$0$}
\rput[t](1.000000,-8.121718){$1$}
\rput[t](2.000000,-8.121718){$2$}
\rput[t](3.000000,-8.121718){$3$}
\rput[t](4.000000,-8.121718){$4$}
\rput[t](5.000000,-8.121718){$5$}
\rput[t](6.000000,-8.121718){$6$}
\rput[r](-0.072000,-8.000000){$10^{-8}$}
\rput[r](-0.072000,-6.000000){$10^{-6}$}
\rput[r](-0.072000,-4.000000){$10^{-4}$}
\rput[r](-0.072000,-2.000000){$10^{-2}$}
\rput[r](-0.072000,0.000000){$10^{0}$}
} 

\psframe[linewidth=\AxesLineWidth,dimen=middle](0.000000,-8.000000)(6.000000,0.000000)

{ \small 
\rput[b](3.000000,-9.3){
\begin{tabular}{c}
$\tau$\\
\end{tabular}
}

\rput[t]{90}(-1.4,-4.000000){
\begin{tabular}{c}
relative frequency\\
\end{tabular}
}
} 

\newrgbcolor{color39.0105}{1  0  0}
\psline[plotstyle=line,linejoin=1,linestyle=solid,linewidth=\LineWidth,linecolor=color39.0105]
(0.050000,-2.704110)(0.150000,-1.754187)(0.250000,-1.319130)(0.350000,-1.039834)(0.450000,-0.838825)
(0.550000,-0.686124)(0.650000,-0.566943)(0.750000,-0.472887)(0.850000,-0.398732)(0.950000,-0.341003)
(1.050000,-0.297271)(1.150000,-0.265774)(1.250000,-0.245190)(1.350000,-0.234503)(1.450000,-0.232914)
(1.550000,-0.239787)(1.650000,-0.254603)(1.750000,-0.276935)(1.850000,-0.306427)(1.950000,-0.342782)
(2.050000,-0.385743)(2.150000,-0.435094)(2.250000,-0.490646)(2.350000,-0.552236)(2.450000,-0.619720)
(2.550000,-0.692972)(2.650000,-0.771880)(2.750000,-0.856347)(2.850000,-0.946283)(2.950000,-1.041609)
(3.050000,-1.142253)(3.150000,-1.248152)(3.250000,-1.359247)(3.350000,-1.475484)(3.450000,-1.596816)
(3.550000,-1.723197)(3.650000,-1.854589)(3.750000,-1.990952)(3.850000,-2.132253)(3.950000,-2.278461)
(4.050000,-2.429545)(4.150000,-2.585480)(4.250000,-2.746238)(4.350000,-2.911798)(4.450000,-3.082137)
(4.550000,-3.257234)(4.650000,-3.437071)(4.750000,-3.621630)(4.850000,-3.810895)(4.950000,-4.004848)
(5.050000,-4.203476)(5.150000,-4.406765)(5.250000,-4.614701)(5.350000,-4.827272)(5.450000,-5.044467)
(5.550000,-5.266275)(5.650000,-5.492684)(5.750000,-5.723686)(5.850000,-5.959270)(5.950000,-6.199428)

\newrgbcolor{color40.01}{0  0  1}
\psline[plotstyle=line,linejoin=1,showpoints=true,dotstyle=*,dotsize=\MarkerSize,linestyle=solid,linewidth=\LineWidth,linecolor=color40.01]
(0.050000,-2.552842)(0.150000,-1.677781)(0.250000,-1.281498)(0.350000,-1.046724)(0.450000,-0.820736)
(0.550000,-0.689944)(0.650000,-0.559091)(0.750000,-0.459170)(0.850000,-0.384260)(0.950000,-0.321664)
(1.050000,-0.297483)(1.150000,-0.269864)(1.250000,-0.243973)(1.350000,-0.238523)(1.450000,-0.230327)
(1.550000,-0.241542)(1.650000,-0.262330)(1.750000,-0.285586)(1.850000,-0.309538)(1.950000,-0.360613)
(2.050000,-0.384681)(2.150000,-0.438899)(2.250000,-0.494850)(2.350000,-0.555331)(2.450000,-0.628563)
(2.550000,-0.692718)(2.650000,-0.760701)(2.750000,-0.856985)(2.850000,-0.965773)(2.950000,-1.069560)
(3.050000,-1.127261)(3.150000,-1.248721)(3.250000,-1.338187)(3.350000,-1.473661)(3.450000,-1.562249)
(3.550000,-1.709965)(3.650000,-1.850781)(3.750000,-2.075721)(3.850000,-2.142668)(3.950000,-2.292430)
(4.050000,-2.309804)(4.150000,-2.537602)(4.250000,-2.619789)(4.350000,-3.221849)(4.450000,-3.221849)
(4.550000,-3.301030)(4.650000,-3.301030)(4.750000,-4.000000)(4.850000,-3.698970)(4.950000,-3.698970)
(5.050000,-4.000000)
\psline[plotstyle=line,linejoin=1,showpoints=true,dotstyle=*,dotsize=\MarkerSize,linestyle=solid,linewidth=\LineWidth,linecolor=color40.01]
(5.250000,-4.000000)(5.350000,-4.000000)

{ \small 
\rput[bl](0.144000,-7.756564){%
\psframebox[framesep=0pt,linewidth=\AxesLineWidth]{\psframebox*{\begin{tabular}{l}
\Rnode{a1}{\hspace*{0.0ex}} \hspace*{0.7cm} \Rnode{a2}{~~Maxwellian} \\
\Rnode{a3}{\hspace*{0.0ex}} \hspace*{0.7cm} \Rnode{a4}{~~Simulation} \\
\end{tabular}}
\ncline[linestyle=solid,linewidth=\LineWidth,linecolor=color39.0105]{a1}{a2}
\ncline[linestyle=solid,linewidth=\LineWidth,linecolor=color40.01]{a3}{a4} \ncput{\psdot[dotstyle=*,dotsize=\MarkerSize,linecolor=color40.01]}
}%
}%
} 

\end{pspicture}%
       \label{fig:greedy_pdf_center}
     }
     \subfigure[PDF at $\omega = \frac{\pi}{4}$]{
%
%
%
\providelength{\AxesLineWidth}       \setlength{\AxesLineWidth}{0.5pt}%
\providelength{\GridLineWidth}       \setlength{\GridLineWidth}{0.4pt}%
\providelength{\GridLineDotSep}      \setlength{\GridLineDotSep}{0.4pt}%
\providelength{\MinorGridLineWidth}  \setlength{\MinorGridLineWidth}{0.4pt}%
\providelength{\MinorGridLineDotSep} \setlength{\MinorGridLineDotSep}{0.8pt}%
\providelength{\plotwidth}           \setlength{\plotwidth}{6cm}
\providelength{\LineWidth}           \setlength{\LineWidth}{0.7pt}%
\providelength{\MarkerSize}          \setlength{\MarkerSize}{2pt}%
\newrgbcolor{GridColor}{0.8 0.8 0.8}%
%
\psset{xunit=0.147318\plotwidth,yunit=0.087144\plotwidth}%
\begin{pspicture}(-0.718894,-8.912281)(6.069124,0.210526)%


\psline[linestyle=dashed,dash=2pt 3pt,dotsep=\GridLineDotSep,linewidth=\GridLineWidth,linecolor=GridColor](0.000000,-8.000000)(0.000000,0.000000)
\psline[linestyle=dashed,dash=2pt 3pt,dotsep=\GridLineDotSep,linewidth=\GridLineWidth,linecolor=GridColor](1.000000,-8.000000)(1.000000,0.000000)
\psline[linestyle=dashed,dash=2pt 3pt,dotsep=\GridLineDotSep,linewidth=\GridLineWidth,linecolor=GridColor](2.000000,-8.000000)(2.000000,0.000000)
\psline[linestyle=dashed,dash=2pt 3pt,dotsep=\GridLineDotSep,linewidth=\GridLineWidth,linecolor=GridColor](3.000000,-8.000000)(3.000000,0.000000)
\psline[linestyle=dashed,dash=2pt 3pt,dotsep=\GridLineDotSep,linewidth=\GridLineWidth,linecolor=GridColor](4.000000,-8.000000)(4.000000,0.000000)
\psline[linestyle=dashed,dash=2pt 3pt,dotsep=\GridLineDotSep,linewidth=\GridLineWidth,linecolor=GridColor](5.000000,-8.000000)(5.000000,0.000000)
\psline[linestyle=dashed,dash=2pt 3pt,dotsep=\GridLineDotSep,linewidth=\GridLineWidth,linecolor=GridColor](6.000000,-8.000000)(6.000000,0.000000)
\psline[linestyle=dashed,dash=2pt 3pt,dotsep=\GridLineDotSep,linewidth=\GridLineWidth,linecolor=GridColor](0.000000,-8.000000)(6.000000,-8.000000)
\psline[linestyle=dashed,dash=2pt 3pt,dotsep=\GridLineDotSep,linewidth=\GridLineWidth,linecolor=GridColor](0.000000,-6.000000)(6.000000,-6.000000)
\psline[linestyle=dashed,dash=2pt 3pt,dotsep=\GridLineDotSep,linewidth=\GridLineWidth,linecolor=GridColor](0.000000,-4.000000)(6.000000,-4.000000)
\psline[linestyle=dashed,dash=2pt 3pt,dotsep=\GridLineDotSep,linewidth=\GridLineWidth,linecolor=GridColor](0.000000,-2.000000)(6.000000,-2.000000)
\psline[linestyle=dashed,dash=2pt 3pt,dotsep=\GridLineDotSep,linewidth=\GridLineWidth,linecolor=GridColor](0.000000,0.000000)(6.000000,0.000000)


\psline[linewidth=\AxesLineWidth,linecolor=GridColor](0.000000,-8.000000)(0.000000,-7.878282)
\psline[linewidth=\AxesLineWidth,linecolor=GridColor](1.000000,-8.000000)(1.000000,-7.878282)
\psline[linewidth=\AxesLineWidth,linecolor=GridColor](2.000000,-8.000000)(2.000000,-7.878282)
\psline[linewidth=\AxesLineWidth,linecolor=GridColor](3.000000,-8.000000)(3.000000,-7.878282)
\psline[linewidth=\AxesLineWidth,linecolor=GridColor](4.000000,-8.000000)(4.000000,-7.878282)
\psline[linewidth=\AxesLineWidth,linecolor=GridColor](5.000000,-8.000000)(5.000000,-7.878282)
\psline[linewidth=\AxesLineWidth,linecolor=GridColor](6.000000,-8.000000)(6.000000,-7.878282)
\psline[linewidth=\AxesLineWidth,linecolor=GridColor](0.000000,-8.000000)(0.072000,-8.000000)
\psline[linewidth=\AxesLineWidth,linecolor=GridColor](0.000000,-6.000000)(0.072000,-6.000000)
\psline[linewidth=\AxesLineWidth,linecolor=GridColor](0.000000,-4.000000)(0.072000,-4.000000)
\psline[linewidth=\AxesLineWidth,linecolor=GridColor](0.000000,-2.000000)(0.072000,-2.000000)
\psline[linewidth=\AxesLineWidth,linecolor=GridColor](0.000000,0.000000)(0.072000,0.000000)


{ \footnotesize 
\rput[t](0.000000,-8.121718){$0$}
\rput[t](1.000000,-8.121718){$1$}
\rput[t](2.000000,-8.121718){$2$}
\rput[t](3.000000,-8.121718){$3$}
\rput[t](4.000000,-8.121718){$4$}
\rput[t](5.000000,-8.121718){$5$}
\rput[t](6.000000,-8.121718){$6$}
\rput[r](-0.072000,-8.000000){$10^{-8}$}
\rput[r](-0.072000,-6.000000){$10^{-6}$}
\rput[r](-0.072000,-4.000000){$10^{-4}$}
\rput[r](-0.072000,-2.000000){$10^{-2}$}
\rput[r](-0.072000,0.000000){$10^{0}$}
} 

\psframe[linewidth=\AxesLineWidth,dimen=middle](0.000000,-8.000000)(6.000000,0.000000)

{ \small 
\rput[b](3.000000,-9.3){
\begin{tabular}{c}
$\tau$\\
\end{tabular}
}

\rput[t]{90}(-1.4,-4.000000){
\begin{tabular}{c}
relative frequency\\
\end{tabular}
}
} 

\newrgbcolor{color46.0114}{1  0  0}
\psline[plotstyle=line,linejoin=1,linestyle=solid,linewidth=\LineWidth,linecolor=color46.0114]
(0.050000,-2.704110)(0.150000,-1.754187)(0.250000,-1.319130)(0.350000,-1.039834)(0.450000,-0.838825)
(0.550000,-0.686124)(0.650000,-0.566943)(0.750000,-0.472887)(0.850000,-0.398732)(0.950000,-0.341003)
(1.050000,-0.297271)(1.150000,-0.265774)(1.250000,-0.245190)(1.350000,-0.234503)(1.450000,-0.232914)
(1.550000,-0.239787)(1.650000,-0.254603)(1.750000,-0.276935)(1.850000,-0.306427)(1.950000,-0.342782)
(2.050000,-0.385743)(2.150000,-0.435094)(2.250000,-0.490646)(2.350000,-0.552236)(2.450000,-0.619720)
(2.550000,-0.692972)(2.650000,-0.771880)(2.750000,-0.856347)(2.850000,-0.946283)(2.950000,-1.041609)
(3.050000,-1.142253)(3.150000,-1.248152)(3.250000,-1.359247)(3.350000,-1.475484)(3.450000,-1.596816)
(3.550000,-1.723197)(3.650000,-1.854589)(3.750000,-1.990952)(3.850000,-2.132253)(3.950000,-2.278461)
(4.050000,-2.429545)(4.150000,-2.585480)(4.250000,-2.746238)(4.350000,-2.911798)(4.450000,-3.082137)
(4.550000,-3.257234)(4.650000,-3.437071)(4.750000,-3.621630)(4.850000,-3.810895)(4.950000,-4.004848)
(5.050000,-4.203476)(5.150000,-4.406765)(5.250000,-4.614701)(5.350000,-4.827272)(5.450000,-5.044467)
(5.550000,-5.266275)(5.650000,-5.492684)(5.750000,-5.723686)(5.850000,-5.959270)(5.950000,-6.199428)

\newrgbcolor{color47.0109}{0  0  1}
\psline[plotstyle=line,linejoin=1,showpoints=true,dotstyle=*,dotsize=\MarkerSize,linestyle=solid,linewidth=\LineWidth,linecolor=color47.0109]
(0.050000,-2.638272)(0.150000,-1.823909)(0.250000,-1.366532)(0.350000,-1.085657)(0.450000,-0.898597)
(0.550000,-0.731656)(0.650000,-0.590405)(0.750000,-0.516270)(0.850000,-0.436045)(0.950000,-0.371100)
(1.050000,-0.333669)(1.150000,-0.283913)(1.250000,-0.264002)(1.350000,-0.243516)(1.450000,-0.243212)
(1.550000,-0.236497)(1.650000,-0.260744)(1.750000,-0.269541)(1.850000,-0.287771)(1.950000,-0.326518)
(2.050000,-0.367543)(2.150000,-0.417255)(2.250000,-0.462181)(2.350000,-0.519706)(2.450000,-0.597911)
(2.550000,-0.660946)(2.650000,-0.732594)(2.750000,-0.792096)(2.850000,-0.879426)(2.950000,-1.007889)
(3.050000,-1.077794)(3.150000,-1.200659)(3.250000,-1.299296)(3.350000,-1.387216)(3.450000,-1.508638)
(3.550000,-1.540608)(3.650000,-1.754487)(3.750000,-1.841638)(3.850000,-2.017729)(3.950000,-2.154902)
(4.050000,-2.327902)(4.150000,-2.481486)(4.250000,-2.638272)(4.350000,-2.677781)(4.450000,-2.744727)
(4.550000,-3.154902)(4.650000,-3.301030)(4.750000,-4.000000)(4.850000,-3.522879)(4.950000,-3.522879)
\psline[plotstyle=line,linejoin=1,showpoints=true,dotstyle=*,dotsize=\MarkerSize,linestyle=solid,linewidth=\LineWidth,linecolor=color47.0109]
(5.150000,-4.000000)(5.250000,-4.000000)
\psline[plotstyle=line,linejoin=1,showpoints=true,dotstyle=*,dotsize=\MarkerSize,linestyle=solid,linewidth=\LineWidth,linecolor=color47.0109]
(5.550000,-4.000000)(5.550000,-4.000000)

{ \small 
\rput[bl](0.144000,-7.756564){%
\psframebox[framesep=0pt,linewidth=\AxesLineWidth]{\psframebox*{\begin{tabular}{l}
\Rnode{a1}{\hspace*{0.0ex}} \hspace*{0.7cm} \Rnode{a2}{~~Maxwellian} \\
\Rnode{a3}{\hspace*{0.0ex}} \hspace*{0.7cm} \Rnode{a4}{~~Simulation} \\
\end{tabular}}
\ncline[linestyle=solid,linewidth=\LineWidth,linecolor=color46.0114]{a1}{a2}
\ncline[linestyle=solid,linewidth=\LineWidth,linecolor=color47.0109]{a3}{a4} \ncput{\psdot[dotstyle=*,dotsize=\MarkerSize,linecolor=color47.0109]}
}%
}%
} 

\end{pspicture}%
       \label{fig:greedy_pdf_corner}
     }
  \end{center}
  \caption{PDF of the DGD at the center frequency ($\omega = 0$) and the corner frequency ($\omega = \frac{\pi}{4}$) for the greedy approximation method. Solid curves represent the Maxwellian with mean 1.6.}
  \label{fig:greedy_dgd_pdfs}
\end{figure*}

At the heart of the third method we propose for the discrete time PMD emulation parameter sampling problem lies a greedy iterative transfer function approximation algorithm \cite{tkacenko2006iterative}. This algorithm takes a general matrix transfer function as input and tries to approximate it with a paraunitary FIR filter. This is achieved by iteratively optimizing each section of the filter. 
The GTFA method first defines a distance measure in the space of discrete time transfer functions as a mean-squared weighted Frobenius norm,

\begin{equation}
  \xi \triangleq \frac{1}{2 \pi} \int \limits_0^{2\pi} W(\omega)\left\|\mathbf{D}(e^{j\omega}) - \mathbf{H}(e^{j\omega})\right\|_F^2 \mathrm{d}\omega \quad,
  \label{eqn:dt_metric}
\end{equation}

\noindent where, for our purposes, $\mathbf{D}(e^{j\omega})$ is the transfer function of a real fiber constructed with a high number of birefringent sections (full model) and $\mathbf{H}(e^{j\omega})$ is the transfer function of the paraunitary FIR filter. $W(\omega)$ is a weighting function which is set  to identity in the frequency range of interest and zero otherwise. This distance is then iteratively minimized by handling the unitary matrix $\mathbf{R}$ and each degree-one section $\mathbf{H}_i$ in equation (\ref{eqn:ll_degreeN}) separately.

The optimization of $\mathbf{R}$ boils down to maximizing $\Re\{\text{tr}(\mathbf{R}^*\mathbf{A})\}$, with

\begin{equation}
  \mathbf{A} = \frac{1}{2 \pi} \int \limits_{0}^{2\pi} W(\omega) \mathbf{V}^*(e^{j\omega}) \mathbf{D}(e^{j\omega}) \mathrm{d}\omega
  \label{eqn:optimize_R}
\end{equation}

\noindent and $\mathbf{V}(e^{j\omega}) = \prod_{i=N}^{1} \mathbf{H}_i(e^{j\omega})$.
This expression can be optimized elegantly by the closest unitary matrix to $\mathbf{A}$ which in turn can be computed via the SVD of $\mathbf{A}$ \cite{horn1990matrix}. 
Similarly the optimization for individual $\mathbf{v}_i$ in equation (\ref{eqn:ll_degree1}) is achieved by minimizing the quadratic form $\mathbf{v}_i^* (\mathbf{B} + \mathbf{B}^*) \mathbf{v}_i = \mathbf{v}_i^* \mathbf{Q} \mathbf{v}_i$ with

\begin{equation*}
  \mathbf{B} =  \frac{1}{2 \pi} \int \limits_{0}^{2\pi} W(\omega) (1-e^{-j\omega})\mathcal{R}_i(e^{j\omega}) \mathbf{R} \mathbf{D}^*(e^{j\omega}) \mathcal{L}_i(e^{j\omega}) \mathrm{d}\omega.
  \label{eqn:optimiza_vi}
\end{equation*}
Here, $\mathcal{R}_i$ and $\mathcal{L}_i$ denote the right and left factors of $\mathbf{V}$ respectively, such that $\mathbf{V}(e^{j\omega}) = \mathcal{L}_i(e^{j\omega})\mathbf{H}_i(e^{j\omega})\mathcal{R}_i(e^{j\omega})$. 
The minimization of this expression is achieved by setting $\mathbf{v}_i$ equal to the corresponding normalized eigenvector of the smallest eigenvalue of $\mathbf{B}$. 
It is shown in \cite{tkacenko2006iterative} that with every such iteration the error term in (\ref{eqn:dt_metric}) is reduced. 
Hence, the algorithm moves forward by optimizing each section and the paraunitary matrix iteratively.

The statistical behavior of this method is demonstrated in Figures  \ref{fig:greedy_mean_corr} and \ref{fig:greedy_dgd_pdfs}. It is obvious that the GTFA method outperforms the previous two strategies at the cost of increased computational complexity.


\section{Summary and Conclusion}
We have presented discrete-time PMD models that make use of a paraunitary FIR filter structure to be used in optical coherent receivers. These models can be incorporated into custom chips or off-line data processing devices present in coherent receivers as DSP based PMD emulators for built-in testing.
The paraunitary FIR filter structure constitutes a good candidate for PMD emulation not only because of its losslessness property but also for its cascaded nature that enables one to adjust the complexity of the filter. Using three different approaches, we have addressed the question of how the parameters of an ensemble of such filters have to be chosen for them to capture the statistical behavior of a real PMD channel.
Using theoretical and simulation results, we have shown that the cascaded sampling method can be employed for systems which suffer mainly from first-order effects of PMD. For higher order effects, the compensated MCMC algorithm can be used which provides a good approximation for the mean and the autocorrelation values of the PMD vector. The final approach approximates the transfer function of an optical fiber in the frequency range of interest and hence provides the best results in terms of desired statistics.
The first two are lower complexity methods which can be implemented in real-time applications while the third one is better suited for off-line signal processing because of its iterative nature.
The choice of appropriate method depends on the trade-off between computational cost and the statistical accuracy.


\end{document}